\documentclass[fleqn,usenatbib]{mnras}

\usepackage[T1]{fontenc}
\usepackage{ae,aecompl}

\usepackage{graphicx}	
\usepackage{amsmath}	
\usepackage{amssymb}	
\usepackage{nccmath}
\usepackage{isotope}
\usepackage{float}
\usepackage{caption}
\usepackage{subcaption}
\usepackage{multicol}
\usepackage[normalem]{ulem}
\usepackage{array}
\usepackage[table]{xcolor}
\usepackage{longtable}
\usepackage{threeparttable}
\usepackage{soul}

\newcommand{\msun} {\mathrm{M}_\odot}
\newcommand{\mstar} {\mathrm{M}_*}

\newcommand{\rsun} {\mathrm{R}_\odot}
\newcommand{\rstar} {\mathrm{R}_*}

\newcommand{\lsun} {\mathrm{L}_\odot}

\newcommand{\ledd} {L_{\mathrm{Edd}}}
\newcommand{\tkh} {\tau_{\mathrm{KH}}}
\newcommand{\feh} {\mathrm{Fe}/\mathrm{H}}

\newcommand{\kms} {\mathrm{kms}^{-1}}
\newcommand{\gs} {\mathrm{gs}^{-1}}
\newcommand{\yr} {\mathrm{a}^{-1}}

\newcommand{\dex} {\mathrm{dex}}

\newcommand{\teff} {T_{\mathrm{eff}}}
\newcommand{\amlt} {\alpha_{\mathrm{MLT}}}

\newcommand{\omegac} {\Omega_{\mathrm{crit}}}

\newcommand{\oomegac} {\Omega / \Omega_{\mathrm{crit}}}
\newcommand{\omegaini} {\omega_{\mathrm{0}}}
\newcommand{\Kelvin} {\mathrm{K}}
\newcommand{\Gauss} {\mathrm{G}}
\newcommand{\gsurf} {\mathrm{log} \, g}
\newcommand{\boltzmann} {k_{\mathrm{B}}}
\newcommand{\Beq} {B_{\mathrm{eq}}}
\newcommand{\massH} {m_{\mathrm{H}}}
\newcommand{\prot} {P_{\mathrm{rot}}}
\newcommand{\turnover} {\tau_{\mathrm{conv}}}
\newcommand{\fmin} {f_{\mathrm{min}}}
\newcommand{\fmax} {f_{\mathrm{max}}}
\newcommand{\fstar} {f_{\mathrm{*}}}
\newcommand{\fsun} {f_{\mathrm{\odot}}}
\newcommand{\rossby}{Ro}

\newcommand{\rossbysun} {Ro_\odot}
\newcommand{\mwind} {\Dot{M}_{\mathrm{wind}}}
\newcommand{\torquewind} {\Gamma_{\mathrm{wind}}}
\newcommand{\ralfven} {R_{\mathrm{A}}}
\newcommand{\vesc} {\nu_{\mathrm{esc}}}
\newcommand{\Bp} {B_{\mathrm{p}}}

\newcommand{\tli}{T_{\mathrm{Li}}}

\title[Rotation, magnetic braking \& Li abundances]{Variable magnetic field and adaptive mixing-length: reproducing Li abundances and constraining rotational evolution of solar-type stars in clusters}

\author[R. Caballero Navarro et al.]{
R. Caballero Navarro,$^{1}$\thanks{E-mail: rcaballeron@correo.ugr.es}
A. Garc\'ia Hern\'andez,$^{1}$\thanks{E-mail: agh@ugr.es}
J.~C. Su\'arez$^{1}$\thanks{E-mail: jcsuarez@ugr.es}
\\
$^{1}$Dept. Theoretical Physics and Cosmology, University of Granada (UGR), 18071, Granada, Spain\\
}

\date{Accepted 2024 XXX XX. Received 2024 XXXX XX; in original form 2024 XXXX XX}

\pubyear{2024}
\usepackage{newtxtext,newtxmath}
\begin{document}
\label{firstpage}
\pagerange{\pageref{firstpage}--\pageref{lastpage}}
\maketitle

\begin{abstract}
Investigating the apparent anomalies in lithium (Li) surface abundance observed in the Sun and young stellar globular clusters within contemporary astrophysical contexts holds significant promise for advancing our understanding of the mechanisms influencing Li depletion throughout stellar evolution. This study delves into the intricate interplay between rotational mixing and rotational hydrostatic effects in pre-main-sequence (PMS) and main-sequence (MS) solar-type stars by employing grids of rotating models. We implement a novel approach in which both the magnetic field strength (B) and the mixing-length parameter ($\amlt$) vary dynamically with stellar parameters, following prescriptions based on dynamo theory and 3D atmosphere calibrations. This avoids fixed values and aims to reduce free parameters while capturing key physical variability.\par

Our models reproduce the observed Li abundance of Sun-like stars (A(Li) $\approx$ 1.12 dex) consistent with the present-day solar value (1.1 $\pm$ 0.1 dex) and yield qualitatively consistent rotational spin-down trends across PMS and MS phases. However, at the solar age (4.57 Gyr), the same models over-predict the equatorial rotation rate (v $\approx$ 4.72 $\kms$ vs. 2.0 $\kms$) and the mean surface magnetic field (B $\approx$ 36.9 G vs. 1 G). These discrepancies reflect the omission of additional angular momentum loss mechanisms—such as disk locking and internal magnetic coupling—and possible oversimplifications in magnetic saturation physics. While the adaptive $\amlt$ converges to the solar-calibrated value ([1.76, 1.78]) at the present age, its variability during earlier phases significantly influences Li depletion.\par

To validate our models, we compare predictions with observational data from 64 open clusters obtained through the Gaia-ESO Survey (GES), sampling a wide range of ages. The results demonstrate that incorporating time-dependent B and $\amlt$ improves Li predictions and captures rotational evolution trends, but cannot yet reproduce the present-day solar rotation and magnetic flux without additional physics. We discuss these limitations and outline future work to integrate disk locking and internal angular momentum transport for a more complete model of solar-type stars.\par

\end{abstract}




\section{Introduction} \label{sec_intro}
Throughout the star's evolution, directly observable parameters such as its rotational velocity, effective temperature, surface gravity, or abundance of chemical elements undergo variations. These parameters exert an influence, either directly or indirectly, on the angular momentum loss (AML) caused by the star's magnetic field. In addition, we encounter stellar convection motions, which in most stellar evolution models, are described by the mixing-length theory (MLT). MLT makes use of the mixing length scale which is proportional to the local pressure scale height, and to the mixing-length parameter ($\amlt$). This parameter, $\amlt$, is considered a fixed free parameter and must be determined by comparing the stellar models with some reference, in most cases our Sun. Therefore, assuming that the magnetic field strength ($B$) and $\amlt$ remain constant throughout the stellar evolution is a necessary simplification in some scenarios but does not correspond with reality.\par

There are different mechanisms that affect the AML in solar-type stars, e.g. tidal forces in binary systems or stellar winds. In close binary systems, angular momentum and kinetic energy can be exchanged between the two stars and their orbits through tidal forces \citep{Li2018}. In the latter case, the magnetically-coupled stellar winds efficiently remove angular momentum,  leading to the spin-down of stars with age. This is due to the braking force exerted by stellar winds, which results in the eventual loss of angular momentum \citep{UdDoula2002}.\par 

In recent years, the rotational evolution of low- and intermediate-mass stars has been studied thanks to several observational campaigns \citep{Hartman2010, Gallet2013, Bouvier2016}. Some specific investigations, such as those by \citet{Ud-Doula2008, Cranmer2011, Jackson2014, Gallet2015, Amard2016}, focus on modeling the effects of magnetic braking (MB) and its coupling with stellar winds on the rotational evolution of stars. The AM transport mechanism in these types of stars is driven by surface removal caused by MB. These models rely on a number of parameters, including the magnetic field strength B, which is not assigned fixed values but rather calculated using various theoretical and observational frameworks. These approaches are not sufficient to explain the observational data, and interaction with other processes in the angular velocity evolution of low-mass stars is necessary: e.g., the distribution of AML in the interior of the star along the pre-main sequence (PMS) and main sequence (MS) \citep{Charbonnel2005, Eggenberger2008, Eggenberger2009, Caballero2020}, and the star - protostellar disc interaction during the PMS \citep{Bouvier2008,Gallet2013,Eggenberger2012,Zanni2012,Bouvier2016}.\par

The impact of rotation on both PMS and Li depletion in solar-type stars has been extensively debated \citep{Pinsonneault1997,Jeffries2004,Somers2014}, with recent revisions based on more accurate measurements \citep{Gallet2013, Bouvier2016, Bouvier2018, Franciosini2022}. Other lines of research are those that point to the possible correlation between the presence of planets \citep{Israelian2009, DelgadoMena2014, Mishenina2016}, as well as between the presence of sunspots and their influence on the Li variability \citep{Jackson2014, Franciosini2022}.\par

Li is destroyed in stellar envelopes when the temperature at the base of the convection zone (BCZ) reaches $\tli \approx 2.5 x 10^6\; \mathrm{K}$. Solar-type stars are known to have a convection zone (CZ) that covers much of the stellar radius during the PMS. In this way, the convective movements that occur in its interior manage to drag Li to areas in which the BCZ limit exceeds $\tli$ and Li is burned \citep{Iben1965}. This process slows down as the star approaches the zero-age MS (ZAMS) when the convection zone retreats and moves away from areas where the internal temperature is high enough to destroy the Li. Discrepancies in Li abundances have been detected in young clusters for which there is as yet no coherent explanation \citep[see][and references therein]{Caballero2020}.\par

A number of different efforts to determine the abundances of Li in open clusters and stars at the PMS stage have been made in the past \citep{Randich2000, Randich2003, Sestito2008}. In the last years, the Gaia-ESO Survey (GES) \citep{Gilmore2012,Randich2013,Randich2022} has produced very high-quality spectroscopy records of around 110,000 Milky Way stars, in the field and in open clusters (OC), down to magnitude 19. It has delivered good-quality radial velocities and stellar parameters for a large fraction of its unique target stars. Elemental abundances for up to 31 elements were derived for targets observed with the Fiber Large Array Multi-Element Spectrograph (FLAMES) and the UV-Visual Echelle Spectrograph (UVES). Li abundances are delivered for about 1/3 of the sample.\par

The primary goal of the work we present here is to explore how MB contributes to the evolution of Li in the Sun and other solar-type stars. We aim to develop evolutionary stellar models in which both $B$ and $\amlt$ are free parameters that vary over time based on specific stellar parameters. We provide semi-empirical approximations for the calculation of AML resulting from braking caused by variable parameters $B$ and $\amlt$. Simulation data are compared with various stellar parameters obtained from Gaia and GES, allowing for the analysis, validation, and refinement of our models against a large candidate set. This has been implemented as an extension to Modules for Experiments in Stellar Astrophysics \citep[MESA; ][]{Paxton2011, Paxton2013,Paxton2015, Paxton2018, Paxton2019}.\par

\section{Data} \label{sec_data}
As indicated by its name, GES intends to complement the data on parallaxes and proper motions from the Gaia satellite \citep{Mignard2005} with extremely precise information on radial velocities (RVs), Li, and chemistry in general. The first Gaia data release (GDR1) \citep{Brown2016} dated 2016 and contained information on the first 14 months of operation. Gaia data release 2 (GDR2) \citep{Brown2018}, released in 2018, contained positions, parallaxes, and proper motions for about 1.3 billion sources, together with photometry. The third data release of Gaia, the Gaia early data release 3 (GEDR3) \citep{Brown2021} and the Gaia data release 3 (GDR3) \citep{Brown2022}, convey updated and more precise measures. GES enhances them and in its latest release (DR5.0) includes all astrophysical parameters derived by the Gaia-ESO consortium \citep{Gilmore2022}. Of interest for the present work are:
\begin{itemize}
    \item The radial and projected rotational velocities
    \item Stellar parameters (effective temperature, surface gravity and metallicity)
    \item Abundances of several elements, among them Li
    \item Cluster probability membership
\end{itemize}

The data sets provided by the Gaia mission in GEDR3 and GDR3, as well as those published by GES in DR5.0 form the main sources used in this work. We focus on OCs to search for solar twins and stars close to the ZAMS where the Li depletion occurs.  We put a special  emphasis on those studied in \citet{Bragaglia2022} and \citet{Randich2022}. Our final sample consists of 64 OCs (see Table \ref{tab:oc_full_list}).

The data sets provided by the Gaia mission in GEDR3 and GDR3, as well as those published by GES in DR5.0 form the main sources used in this work. We focus on OCs to search for solar twins and stars close to the ZAMS where the Li depletion occurs. We used the catalogues provided by \citet{Bragaglia2022} and \citet{Randich2022}. Our final sample consists of 64 OCs (see Table \ref{tab:oc_full_list}). 

\begin{table*}
	\centering
	\begin{tabular}{|l l l l || c c c c c c c c c c|} 
		\hline
             & & & & & & & & $\omegaini$ & & & & & \\
		Open Cluster & Age (Ga) & [Fe/H] & $N_*$ & 0.095 & 0.10 & 0.105 & 0.11 & 0.115 & 0.12 & 0.125 & 0.13 & 0.14 & 0.1425\\
		\hline
            Berkeley 21 & 2.138 & -0.21 & 744 & 0 & 0 & 0 & 0 & 0 & 1 & 1 & 1 & 1 & 1\\
            Berkeley 39 & 5.623 & -0.14 & 899 & 1 & 1 & 1 & 1 & 1 & 1 & 2 & 2 & 2 & 2\\
            IC 2602 & 0.036 & -0.06 & 1836 & 0 & 0 & 0 & 0 & 0 & 0 & 1 & 1 & 0 & 0\\
            IC 4665 & 0.033 & 0.01 & 567 & 0 & 0 & 0 & 0 & 1 & 1 & 0 & 0 & 0 & 0\\
            Messier 67 & 3.981 & -0.02 & 131 & 6 & 6 & 6 & 6 & 6 & 6 & 6 & 6 & 6 & 6\\
            NGC 2141 & 1.862 & -0.04 & 853 & 1 & 1 & 1 & 1 & 1 & 1 & 1 & 1 & 1 & 1\\
            NGC 2355 & 1 & -0.13 & 208 & 1 & 1 & 1 & 1 & 1 & 1 & 1 & 1 & 1 & 1\\
            NGC 2420 & 1.698 & -0.15 & 562 & 1 & 1 & 1 & 1 & 1 & 1 & 1 & 1 & 1 & 1\\
            NGC 2425 & 2.399 & -0.13 & 528 & 1 & 1 & 1 & 1 & 1 & 1 & 1 & 1 & 1 & 1\\
            NGC 2451 & 0.035 & -0.08 & 1656 & 0 & 0 & 0 & 0 & 0 & 1 & 1 & 0 & 2 & 1\\
            NGC 2516 & 0.24 & -0.04 & 759 & 0 & 0 & 0 & 1 & 1 & 1 & 3 & 4 & 4 & 3\\
            NGC 3532 & 0.398 & -0.01 & 1145 & 1 & 1 & 0 & 1 & 1 & 1 & 1 & 1 & 0 & 1\\
            NGC 6005 & 1.259 & 0.22 & 560 & 0 & 0 & 0 & 0 & 0 & 0 & 0 & 1 & 1 & 1\\
            NGC 6259 & 0.269 & 0.18 & 494 & 0 & 0 & 0 & 0 & 0 & 0 & 0 & 0 & 0 & 1\\
            NGC 6281 & 0.513 & -0.04 & 320 & 0 & 0 & 0 & 0 & 0 & 0 & 0 & 0 & 1 & 1\\
            NGC 6405 & 0.035 & -0.02 & 701 & 1 & 1 & 1 & 0 & 1 & 3 & 2 & 0 & 0 & 0\\
            NGC 6633 & 0.692 & -0.03 & 1662 & 2 & 2 & 2 & 1 & 1 & 0 & 0 & 0 & 1 & 1\\
            NGC 6709 & 0.191 & -0.02 & 730 & 2 & 1 & 0 & 0 & 0 & 0 & 0 & 0 & 1 & 1\\
            Trumpler 20 & 1.862 & 0.13 & 1213 & 3 & 3 & 3 & 2 & 2 & 2 & 2 & 2 & 2 & 2\\
            \hline
	\end{tabular}
 	\caption{List of selected OCs. For each OC, name, estimated age, metallicity and number of components are listed. Additionally, the different $\omegaini$ used in the different simulations are shown, where $\omegaini = \oomegac$, $\Omega$ is the star angular velocity at stellar surface, and $\omegac$ is the surface velocity at the equator of a rotating star where the centrifugal force balances the Newtonian gravity. For each OC entry, it is informed per $\omegaini$ how many OC components whose stellar parameters, after comparing them with the equivalent ones calculated by the simulations, are selected. The sorted-out OC members are used as a reference in the pictures shown in the following sections.}
  	\label{tab:oc_reduced_list}
\end{table*}

\subsection{The selection of clusters}
GES DR5.0 provides information on 114 324 stars corresponding to OCs \cite[see][for details]{Gilmore2022}. We select those records which satisfy any of the conditions:
\begin{itemize}
    \item GE\_CL: observed by GES, OC programme field
    \item GE\_SD\_OC: observed by GES, standard field OC
    \item AR\_CL: ESO Archive observation, OC programme field
    \item AR\_SD\_OC: Archive observation, standard field OC
\end{itemize}

As a result of this first filtering, the number of selected records is reduced to 43 299. On this subset of data we perform another filtering, this time aimed at selecting those that offer not null information on the following attributes:
\begin{itemize}
    \item Metallicity
    \item Li abundance
    \item GES field identifier
    \item Effective temperature
    \item Surface gravity
    \item Cluster membership probability (>= 0.95)
\end{itemize}

This second filtering reduces the population of stars to be considered to a total of 5 895 records which are distributed among the 64 OCs listed in Table \ref{tab:oc_full_list}.\par

As benchmark data we take a subset of the observations obtained by GES. This subset consists of those stars that belong to OCs, and among them are "hidden" potential twins of our Sun. To identify the latter, an additional screening is performed, this time using the values given by our models during the simulations for metallicity ($[\feh]$), effective temperature ($\teff$), and surface gravity ($\gsurf$). Finally we are left with those stars for which the GES observations have measured values congruent with our model results, from which we obtain their Li abundances.\par

\subsection{The selection of solar analogues}
Once we have pre-selected the stars belonging to the OCs, our interest is focused on identifying those that are most similar to our Sun at different ages, i.e. solar analogues. This brings the question, which stellar parameters should be considered to classify a star as solar analogue? Finding them would help us to figure out the origin of our Sun and and gain information about what conditions prevailed in the OCs in which they were born.\par

Several attempts have already been made to find solar siblings and solar twins. The former refers to stars that formed in the same cluster as the Sun and therefore they should have ages and chemical abundances very similar to those of the Sun \citep[see][and references therein]{Adibekyan2018}. However, according to \citet{Strobel1996} solar twins are considered stars possessing fundamental physical parameters such as $\teff$, $\gsurf$, $\feh$, photometric properties, chemical composition, age, luminosity, rotation, and magnetic fields that are similar, if not identical, to those of the Sun. Our approach is somewhere between the two approaches, since we are interested in locating stars with $1\,\msun$, similar solar metallicity, $\teff$ and $\gsurf$. Concerning $\teff$ and $\gsurf$, we consider them at different evolutionary times depending on the cluster age in which they are located i.e., taken as reference values those produced by our Sun-like models at the cluster age. Locating these solar analogues presents a formidable challenge due to strict criteria and their dispersion throughout the Milky Way.\par 

Commonly, the applied search methods focus on identifying candidates through their kinematic properties, after which their temperatures, metallicities, and chemical abundances are compared with those of the Sun. In our study, we extend the selection criteria by adding the surface gravity and the age of the star. These two new parameters represent a fine-tuning of the search process for twin candidates. We used the age of the OCs reported in Table \ref{tab:oc_reduced_list} and the data obtained from the set of MESA simulations initiated with different $\omegaini$, where $\omegaini = \oomegac$, $\Omega$ is the angular velocity of the star at the stellar surface, and $\omegac$ is the surface velocity at the equator of a rotating star where the centrifugal force balances Newtonian gravity. The age derived from these simulations provides a crucial parameter for our analysis. This information enables us to determine the remaining simulation parameters, which we subsequently use for comparison with the values associated with stars belonging to the short-listed OCs.\par

\begin{table}
	\centering
	\begin{tabular}{ll} 
		\hline
		Stellar Parameter & Selection Interval\\
		\hline
        OC Age & $\pm0.1\%$ \, for age < 1 Ga \\
        & $\pm0.15\%$ \, for age $\geq$ 1 Ga \\
        $\teff$ & $\pm 50.0 \, \Kelvin$\\
        $\gsurf$ & $\pm 0.05 \, \dex$\\
        $\feh$ & $\pm 0.05 \, \dex$\\
		\hline
	\end{tabular}
    \caption{Selection parameters and intervals applied during the triage process over the OC's components.}
    \label{tab:sel_params}
\end{table}

From approximately 900 identified stars, we select those whose $\feh$, $\teff$, and $\gsurf$ are in the correspondent intervals defined in Table \ref{tab:sel_params} and according to the following classification process: based on the values given by our simulations for the parameters listed in Table \ref{tab:sel_params}, we proceed to select from the different OCs those stars that have values within the established ranges. Specifically, for the estimated age of a given cluster, we establish an interval of $\pm$0.1 for young OCs (age < 1 Ga) and $\pm$0.15 for older counterparts (age >= 1 Ga) \citep[refer to][for comprehensive reference values]{Cantat-Gaudin2020}. This age-dependent interval serves as the basis for selecting simulation time steps that align with the lower and upper age limits of each cluster. Once these two time steps are found, the associated limit values for $\teff$, $\gsurf$ and $\feh$ are gathered. With these value ranges, we proceed then to select in each of the OCs those stars that have values for these same parameters within them. In this way, we finally obtain the candidate stars to be compared with the simulation results.\par

\begin{table*}
	\centering
	\begin{tabular}{l l l l l l l} 
		\hline
            ObjectId & $\teff$(K) & $\gsurf$(dex) & FeH(dex) & ALi(dex) & eALi(dex) & Age(Ga)\\
		\hline
            8510914-1157003 & 5843 & 4.42 & -0.03 & 1.8 & 0.08 & 3.981\\ 
            8510991-1146169 & 5797 & 4.4 & -0.01 & 1.65 & 0.08 & 3.981\\ 
            8512177-1144050 & 5818 & 4.46 & 0 & 1.99 & 0.06 & 3.981\\ 
            8513941-1200571 & 5792 & 4.41 & -0.02 & 1.49 & 0.13 & 3.981\\ 
            8515559-1148381 & 5794 & 4.43 & -0.04 & 1.56 & 0.1 & 3.981\\ 
            8520350-1147480 & 5875 & 4.47 & -0.03 & 1.21 & 0.24 & 3.981\\ 
            \hline
	\end{tabular}
 	\caption{M67 - Filter setup: GES star identifier, $\teff$(K) [5787.4892, 5887.6823], $\gsurf$(dex) [4.3979, 4.4977], $\feh$(dex) [-0.05, 0.05], Age(Ga) [3.977, 3.985]}
	\label{tab:oc_m67}
\end{table*}

\begin{table*}
	\centering
	\begin{tabular}{l l l l l l l} 
		\hline
            ObjectId & $\teff$(K) & $\gsurf$(dex) & FeH(dex) & ALi(dex) & eALi(dex) & Age(Ga)\\
		\hline
            7571111-6048156 & 5084 & 4.54 & 0 & 2.29 & 0.08 & 0.24\\ 
            7595031-6044149 & 5108 & 4.52 & -0.05 & 1.92 & 0.08 & 0.24\\ 
            7595280-6032498 & 5093 & 4.49 & -0.03 & 2.29 & 0.08 & 0.24\\ 
            \hline
	\end{tabular}
 	\caption{NGC2516 - Filter setup: GES star identifier, $\teff$(K) [5010.0324, 5111.0095], $\gsurf$(dex) [4.4405, 4.5405], $\feh$(dex) [-0.05, 0.05], Age(Ga) [0.23964, 0.24036]}\label{tab:oc_ngc2516}

\end{table*}

The Tables \ref{tab:oc_m67} and \ref{tab:oc_ngc2516} summarize the subset of components of the OC's M67 and NGC2516 selected during the simulations for models initialised with $\omegaini = 0.14$. Each of the tables lists the GES identifier of the star and $\teff$, $\gsurf$, $\feh$, A(Li) and estimated age ranges used in the simulation.

\section{The modelling} \label{modelling}
\subsection{Modelling the rotational mixing}
Modeling the effects of rotation on the evolution of the star is a fundamental aspect to consider, and the physical framework involved is a 3D process. Its modeling requires large computational power and, in most cases, its simulation is not feasible. One approach that allows us to address this limitation is the layered (\textit{shellular}) \citep{Meynet1997} treatment, which allows us to solve the stellar structure equations in 1D. In this way, the hydrostatic effects of rotation in 1D stellar models can be calculated.\par

There are different stellar evolution codes that support the modeling of the rotation and the effects that they have on the evolution of the star, among which are the instabilities that they trigger, the angular moment transport, or their impact on the mixing effects. Among these codes, we can find Geneva \citep{Eggenberger2008}, YREC \citep{Demarque2008}, STAREVOL \citep{Dumont2021, Amard2022}, and MESA \citep{Paxton2010, Cantiello2014}. All these codes have a number of similarities and some special aspects in their treatment.
In MESA, convection is modeled using the Mixing Length Theory \citep{Paxton2010}, with a number of variations available. While the Tayler instability is widely recognized, the Tayler-Spruit (TS) dynamo loop remains a subject of ongoing discussion, both in theoretical and computational studies \citep{Braithwaite2006, Zahn2007, Mathis2008}. Angular momentum transport in convective regions is managed through the MLT diffusion coefficient, which is large enough to enforce rigid rotation within these zones.\par

We aim to use the most coherent models although it is not our aim to make an exhaustive comparison between all of them. We use as a starting reference the work of \cite{Choi2016}, and by extension MESA as a code for stellar evolution. More specifically we focus our interest on the comparison of results, for lithium abundance and rotation period, that we obtain when we apply in $\amlt$ and variable magnetic field strength.\par

Our models, as shown in Table \ref{tab:phy_mesa}, include rotation during PMS and MS. During the MS stage, the star exhibits a solid body rotation already because the interior of the star is fully convective. Once the star approaches the ZAMS, before entering the MS, instabilities begin to develop that cause the appearance of differential rotation between the radiative core and the convective envelope. We consider the following rotation-induced instabilities: Solberg-Hoiland (SH) instability, Eddington-Sweet (ES) circulation, Goldreich–Schubert–Fricke (GSF) instability, secular shear instability (SSI), and dynamical shear instability (DSI) excluding the Spruit-Taylor (ST) treatment.\par 

The chemical mixing and the transport of angular momentum due to internal magnetic ﬁelds are not included in our models, though this is implemented in MESA following \cite{Heger2005}. Our work does not question the ST dynamo but rather studies a different dynamo prescription and therefore, we disable its effects in our simulations.\par

\subsection{Modelling the mass loss} \label{mod_mass_loss}
A number of theoretical mass-loss schemes exist in the literature. These support different mechanisms and are relevant to particular types of stars and the evolutionary phase they are in. It is worth mentioning the Langer formalism \citep{Langer1998}, which establishes an empirical relation that determines how the stellar mass loss increases as the rotational velocity increases and approaches the critical velocity.\par 

On the other hand, for evolutionary phases in which stellar winds represent the main mechanism by which mass loss occurs, the prescription of Reimers \citep{Reimers1975} is the one used as a reference. It applies mainly to stars in advanced evolutionary phases, such as red giants (RGB) and asymptotic giant branch (AGB) stars. This formalism relates the mass loss rate to the luminosity, radius and mass of the star, and is especially useful for modeling mass loss in cool, luminous stars.\par

It is known that stars with similar initial masses but different mass loss ratios ($\Dot{M}$) will eventually evolve very differently. Ionized particles carried by the solar wind not only contribute to $\Dot{M}$, but also to the loss of
kinetic energy that is deposited in the interstellar medium. In our investigation, we have resorted to a formalism which takes into account the effects of surface angular velocity on mass loss, in particular the approach described in \cite{Paxton2013} which is based on Langer's work \citep{Langer1998}. In this formalism, the surface and critical angular velocities ($\oomegac$), the stellar mass ($M$), the radius ($L$), and the Eddington luminosity ($\ledd$) play a role,\par

\begin{ceqn}
\begin{align}
     \Dot{M}(\Omega) &= \Dot{M}(0)\Bigg(\frac{1}{1-\oomegac}\Bigg)^{0.43} \label{eq:langer}\\
     \omegac^2 &= \Bigg(1 - \frac{L}{\ledd}\Bigg)\frac{GM}{R^3}\\
     \ledd &= \frac{4\pi cGM}{\kappa}
\end{align}
\end{ceqn}

where $c$ is the speed of light and $\kappa$ is the opacity of the star's material.\par

The mass loss saturates as the angular velocity approaches the critical velocity. In this, the minimum value ($\mwind$) between the value obtained according to Langer's formalism and the thermal timescale ($\tkh$) is selected. This is related to the mass loss because it determines how quickly a star can radiate away its energy. A shorter thermal timescale means the star is losing energy more rapidly, which can influence the rate at which it loses mass through stellar winds or other processes.\par

\begin{ceqn}
\begin{equation}
    \mwind = \mathrm{min} \Bigg[\Dot{M}(\Omega), 0.3\frac{M}{\tkh}\Bigg]
\end{equation}
\end{ceqn}

\subsection{Modelling the magnetic braking} \label{mod_mb}
We start by characterizing how the loss of angular momentum occurs in our model. Let us start by enumerating the most relevant aspects and assumptions made in modeling the evolution of rotation, magnetic braking and angular momentum. If we assume a spherical outflow, the torque applied by a magnetically-coupled stellar wind $\torquewind$ is dictated by,
\begin{ceqn}
\begin{equation}
    \torquewind \propto \Omega_* \; \mwind \; \ralfven^{2} \label{eq:mb_torque}
\end{equation}
\end{ceqn}
where $\mwind$ is the mass loss rate, and $\ralfven$ the averaged value of the Alfvén radius.\par

It has been observed that strongly magnetic intermediate-mass stars typically have rotation rates much slower than other stars in their parent population \citep{Mathys2006}. In those stars, the magnetic fields interact with the mass loss, where the Alfv\'{e}n radius plays an important role. $\ralfven$ is defined as the point in which the magnetic field energy density and the kinetic energy density are balanced. In case that $\ralfven$ is greater than the stellar radius, then the wind flow will have to follow the magnetic field. As a consequence, the material leaves the stellar surface with a higher specific AM, as the co-rotation radius has increased and it roughly corresponds to $\ralfven$ which can be expressed as \citep{Matt2012}
\begin{ceqn}
\begin{equation}
    \ralfven = K_1\left[\frac{\Bp^{2}\;R_*^{2}}{\mwind\;\sqrt{K_2^2\vesc^2 + \Omega_*^2R_*^{2}}\ }\right]^{m}R_*  \label{eq:mb_ralfven}
\end{equation}
\end{ceqn}
where $m =0.1675$, $K_1 = 1.30$, and $K_2 = 0.0506$ \citep{Gallet2013}, and $\Bp$ is the mean surface magnetic field and $\vesc$ is the escape velocity defined as:
\begin{ceqn}
\begin{equation}
\vesc = \sqrt{\frac{2\,G\,\mstar}{\rstar}} \label{eq:vesc}
\end{equation}
\end{ceqn}
Finally, we have that the AML can be calculated as follows:
\begin{ceqn}
\begin{equation}
 \Dot{J} = \Omega_* \; \mwind \; \ralfven^{2} \label{eq:j_dot}
\end{equation}
\end{ceqn}

\subsection{Modelling the magnetic intensity} \label{mod_mbi}
Our modeling approach is based on variations of the magnetic field strength throughout a star's evolution, following the formalism introduced by \citet{Gallet2013}. The adopted model employs a dynamo-type prescription, establishing relationships among the star's angular velocity ($\Omega$), effective temperature ($\teff$), magnetic field strength ($B$), and the mass-loss rate ($\mwind$) induced by the stellar wind. We start by enumerating the most relevant aspects and assumptions made in modeling the evolution of magnetic field intensity. \par 

A relevant parameter to characterize the influence of a given magnetic field on the stellar wind is $\Omega$. Assuming that the mean surface magnetic field ($\Bp$) is generated by a dynamo, its strength is proportional to some power, 
\begin{ceqn}
\begin{equation}
    \Bp \propto \Omega_*^b \label{eq:mf_strenght}
\end{equation}
\end{ceqn}

where $\Omega_*$ is the star angular velocity at the stellar surface, and $b$ is the dynamo exponent.\par

Eq. \ref{eq:mf_strenght} illustrates the general dynamo scaling concept, but our implementation follows an alternative one where $\Omega_*$ enters through the Rossby number and filling factor rather than a direct power law. Consequently, an explicit exponent $b$ is not defined, as the dependence is non-linear and mediated by convective turnover time and surface magnetization.\par

According to  \cite{Gallet2013}, $\Bp$ can be defined by

\begin{ceqn}
\begin{equation}
    \Bp = f_*B_* \label{eq:bp}
\end{equation}
\end{ceqn}

where $f_*$ is the filling factor, i.e. the fraction of the star surface which is magnetized, $B_*$ is the magnetic field strength.\par

Following this approach, the magnetic field strength will be proportional to the equipartition magnetic field strength, $\Beq$, as follows:
\begin{ceqn}
\begin{equation}
    B_* \approx 1.13 \Beq \label{eq:mf_bstar}
\end{equation}
\end{ceqn}
where $\Beq$ is defined, and thus related to $\teff$, as:
\begin{ceqn}
\begin{equation}
    \Beq = \sqrt{8\upi P_*} = \sqrt{\frac{8\upi\rho_* \boltzmann \teff}{\mu\massH}}\label{eq:mf_beq}    
\end{equation}
\end{ceqn}
where $P_*$ is the photospheric gas pressure, $\rho_*$ is the photospheric density, $\boltzmann$ is Boltzmann's constant, $\massH$ is the mass of a hydrogen atom, and $\mu$ is the mean atomic weight \citep{Cranmer2011}.

As indicated in \citet{Cranmer2011}, $\mu$ can be estimated using the OPAL plasma equations of state\footnote{\url{https://opalopacity.llnl.gov/EOS_2005/}},
\begin{ceqn}
\begin{equation}
    \mu \approx \frac{7}{4} + \frac{1}{2} \; {\rm tanh}\Bigg(\frac{3500-\teff}{600}\;\Bigg) \label{eq:mf_atom_weight}
\end{equation}
\end{ceqn}
Thus, $B_*$ can be calculated with the stellar parameters at any evolutionary state of a star.\par

On the other hand, \citet{Cranmer2011} found that the rotation period of the star ($\prot$), and therefore $\Omega_*$, has a major influence over $f_*$ than over $B_*$. That is, variations in the star's angular velocity do not significantly alter the magnetic field strength.\par 

On the other hand, it was found that $f_*$ has a strong dependence on the Rossby number ($\rossby$),
\begin{ceqn}
\begin{equation}
    \rossby = \frac{\prot}{\turnover} \label{eq:mf_rossby}
\end{equation}
\end{ceqn}

In order to calculate $\rossby$ of a given star, it is necessary to know the turnover time ($\turnover$),

\begin{ceqn}
\begin{equation}
    \turnover = 314.24\;exp\left[-\Bigg(\frac{\teff}{1952.5}\Bigg)-\Bigg(\frac{\teff}{6250}\Bigg)^{18} \;\right]+0.002 \label{eq:mf_turnover}
\end{equation}
\end{ceqn}
where $\turnover$ is expressed in days and the ﬁt is valid for the
range $3300 \leq \teff \leq 7000$ K. Such a ﬁt ignores
how $\turnover$ may depend on other stellar parameters besides $\teff$ \citep{Cranmer2011}.

Although $\amlt$ variations in MLT would theoretically affect convective velocities and turnover times, we adopt the empirical fit from \citet{Cranmer2011} to compute $\turnover$. This ensures consistency with the rotation–activity formalism and avoids introducing additional free parameters. The Rossby number calculation is therefore decoupled from $\amlt$ changes, following the approach of \citet{Gallet2013}.

\citet{Cranmer2011} provides two different settings for $\fstar$ which define, respectively, the upper ($\fmax$) and lower limits ($\fmin$) for $\Bp$. Each of them attempts to characterize the fill factors associated with magnetic flux tubes. The former is for the closed flux tubes responsible for the magnetically active regions in the stellar surface. Whereas the latter is associated with the open flux tubes associated with regions with less intense magnetic fields. In this way, it is assumed that $\fstar = \fmin$ when no active regions are present on the stellar surface.

\begin{ceqn}
\begin{align}
     \fmin &= \frac{0.5}{(1+(x/0.16)^{2.6})^{1.3}} \label{eq:fmin}\\
     \fmax &= \frac{1}{1+(x/0.31)^{2.5}} \label{eq:fmax}
\end{align}
\end{ceqn}
where x = $\rossby$/$\rossbysun$, and $\rossbysun$ = 1.96.

Finally, we are able to calculate the filling factor for a star by applying the adjustment made by \citet{Gallet2013} which more closely reproduces the average filling factor of the present Sun $\fsun$ ([0.001-0.01], see \cite{Cranmer2011}),
\begin{ceqn}
\begin{equation}
     \fstar = \frac{0.55}{(1+(x/0.16)^{2.3})^{1.22}} \label{eq:fstart}
\end{equation}
\end{ceqn}

\begin{figure}
	\includegraphics[clip,width=\columnwidth]{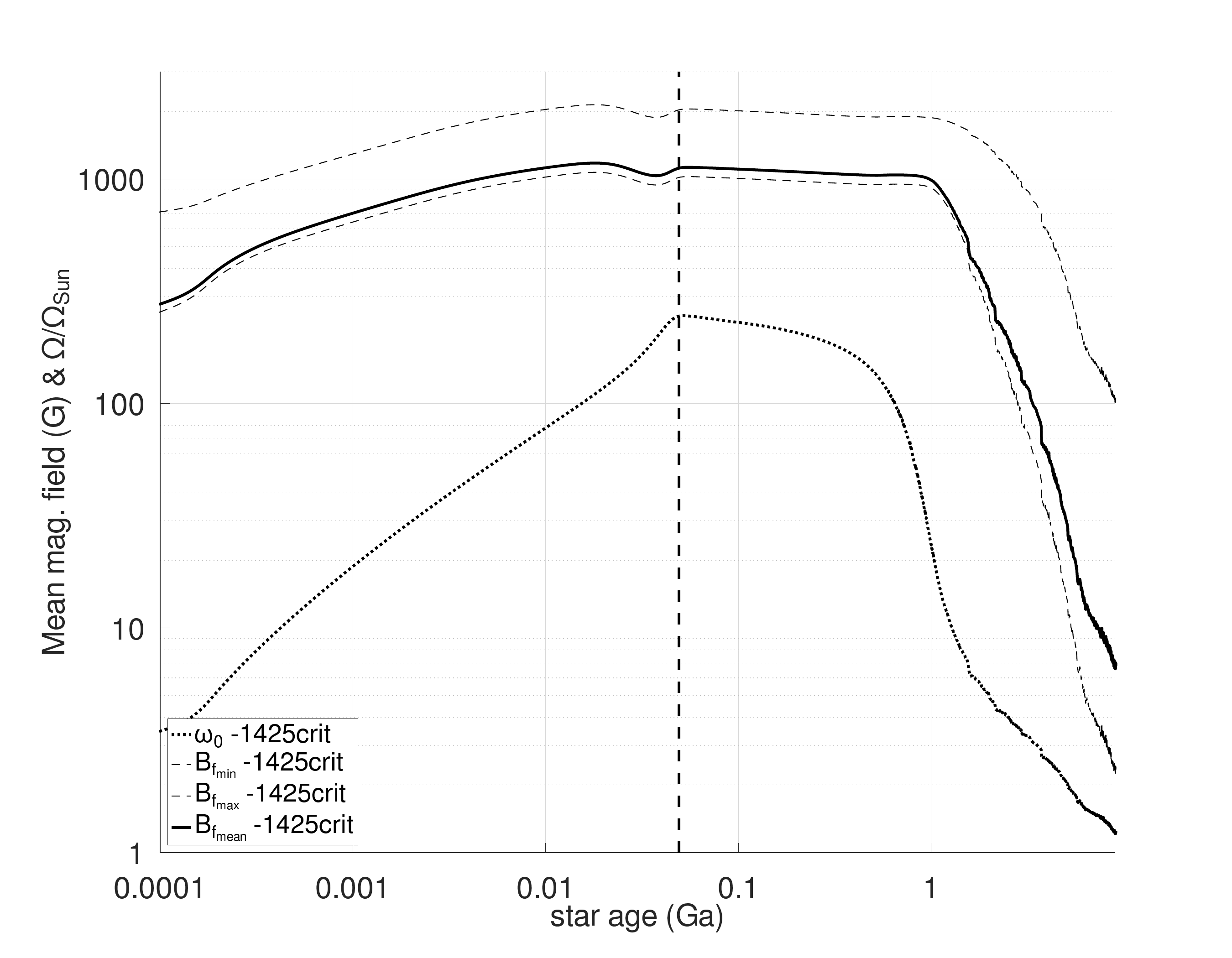}
    \caption{The evolution of magnetic field intensity and its upper and lower limits, as a function of time and $\omegaini$ for a 1 $\msun$ model. The model include rotation with $\omegaini$ = 0.1425. The solid lines represent the magnetic field strength, while the dotted lines represent the angular evolution of the star. The dashed vertical line makes reference to the ZAMS.}
    \label{fig:mag_field_limits_var_vel_g3}
\end{figure}

With this formalism, we are able to calculate $\Bp$ as a time-dependent function of $\rho$, $\teff$, and $\Omega$. The time values for these three variables will be obtained from the different simulations carried out with MESA.\par

As the star undergoes contraction during the pre-main-sequence (PMS) phase towards the Zero-Age Main Sequence (ZAMS), its angular velocity increases, leading to a gradual augmentation in the mean magnetic field. In our analysis, we observed intensities that reached $1.0\,\mathrm{kG}$, as depicted in Figure \ref{fig:mag_field_limits_var_vel_g3}. It then enters a saturation stage (although with a downward trend) up to $1.0$ Ga, and finally declines sharply as a result of the star's slower angular velocity. In contrast,  \citet{Caballero2020} employed a model suited for magnetic fields of non-variable intensity and ranging from a few Gauss ($1.0\,\mathrm{G} \leq B \leq 10.0\,\mathrm{G}$). The significantly higher intensities required in the present formalism render the previously employed model inadequate.\par

This behavior reproduces the qualitative rotation–activity pattern observed in young clusters and the Sun, which is supported by studies of PMS solar-type stars. They report global magnetic field strengths of order kG in classical T Tauri stars and young solar analogues \citep{Johns-Krull2007, Hussain2013, Hill2019}. These findings confirm that strong magnetism is typical in early evolutionary phases and justify the high magnetic intensities predicted by our models during PMS. The absolute field strengths predicted here exceed the present-day solar mean, a limitation discussed in section \ref{sec_3}.\par

Our model follows the following empirical trend: the filling factor and magnetic field strength are computed as functions of $\rossby$ and $\teff$ (Eqs. 14-18), producing strong fields during the PMS when $\rossby$ is small and a gradual decline as rotation slows and $\rossby$ increases. The adopted magnetic field prescription is consistent with the observed rotation–activity correlation in solar-type stars. \citet{Noyes1984} documented that stellar activity scales primarily with the Rossby number, rather than rotation period alone, establishing $\rossby$ as the key dynamo parameter. Many authors (e.g. \citeauthor{Wright2011} \citeyear{Wright2011}) have extended this relation to a large sample of stars, confirming that magnetic activity saturates for fast rotators (low $\rossby$) and declines with increasing $\rossby$ for slower rotators.\par

\subsection{Modelling adaptive MLT $\alpha$} \label{mod_mltalpha}
The mixing-length theory (MLT) introduced by \citet{BohmVit58} has been commonly adopted to model stellar convection in stellar evolutionary codes, including MESA. The most prominent parameter in this theory is the mixing length $l$ which is defined as 
\begin{ceqn}
\begin{equation}
 l = \amlt H_p\label{eq:mixlength}
\end{equation}
\end{ceqn}
where $H_p$ is the pressure scale height, and $\amlt$ is a free parameter that is determined in advance and is normally kept fixed during the stellar evolution simulations.\par

 It became evident that the influence of the free parameter associated with the MLT significantly conditioned the evolution of the Li abundance. Moreover, there are no strong arguments that suggest that the mixing-length parameter is the same in all stars and at all evolutionary phases \citep{Pasetto2014}. In \citet{Caballero2020} the authors showed how an alternative mixing-length (ML) parameterization could produce results in line with the observations. Adjustment of ML over-shooting and $\amlt$ free parameter seems to be required for explaining Li abundances in OC's.\par

In \citet{Sonoi2018}, the authors introduced a method for calibrating $\amlt$ for solar-like stars, in which its value increases with increasing $\gsurf$ or decreasing $\teff$. They calibrated the values of $\amlt$ for several simulated 3D models using the CO$^5$BOLD code \citep{Freytag2011} and transferred them to the 1D models developed by \citet{Ludwig1998}. The parameterization they obtained was: 

\begin{ceqn}
\begin{equation}
 f(x,y) = a_0 + (a_1 + (a_3 + a_5x +a_6y)x + a_4y)x + a_2y\label{eq:alpha_ml}
\end{equation}
\end{ceqn}
where
\begin{ceqn}
\begin{align}
     x &= \frac{\teff-5777}{1000} \label{eq:eq:alpha_x}\\
     y &= \gsurf-4.44 \label{eq:eq:alpha_y}
\end{align}
\end{ceqn}
and $a_i$ are the resulting coefficients of the fitting function for the calibrated $\amlt$ values in the MLT convection model \citep{Sonoi2018}: $a_0=1.790295$, $a_1=-0.14954$, $a_2=0.069574$, $a_3=-0.00829$, $a_4=0.013165$, $a_5=0.080333$, $a_6=-0.03306$. We used this parametrization in our models, applying an adaptive $\amlt$ throughout the stellar evolution.\par

Although the $\amlt$ calibration of \citet{Sonoi2018} is derived from non–rotating 3D hydrodynamical simulations, its application to the present work is valid. First, the calibration is based on $\gsurf$ and $\teff$, which characterise the thermal stratification of the outer convective layers, where the MLT formalism is applied. Even in rotating models, $\gsurf$ and $\teff$ remain the fundamental surface parameters governing convective efficiency, and thus provide a consistent way to evaluate $\amlt$ within the limits of the 1D approach.\par

Second, rotation triggers centrifugal effects on the surface layers leading to changes in the stellar radius. The resulting variations in $\gsurf$ ($\propto R^{-2}$) and $\teff$ (through $L \propto R^2\teff^4$) are considered by the interpolation domain of the $\amlt$ ($\gsurf$, $\teff$) calibration. In practice, we are assuming that the dominant effect on $\amlt$ arises from the thermal structure rather than rotation itself.\par

\subsection{Grid of Models} \label{sec_grid}
\begin{figure}
	\includegraphics[clip,width=\columnwidth]{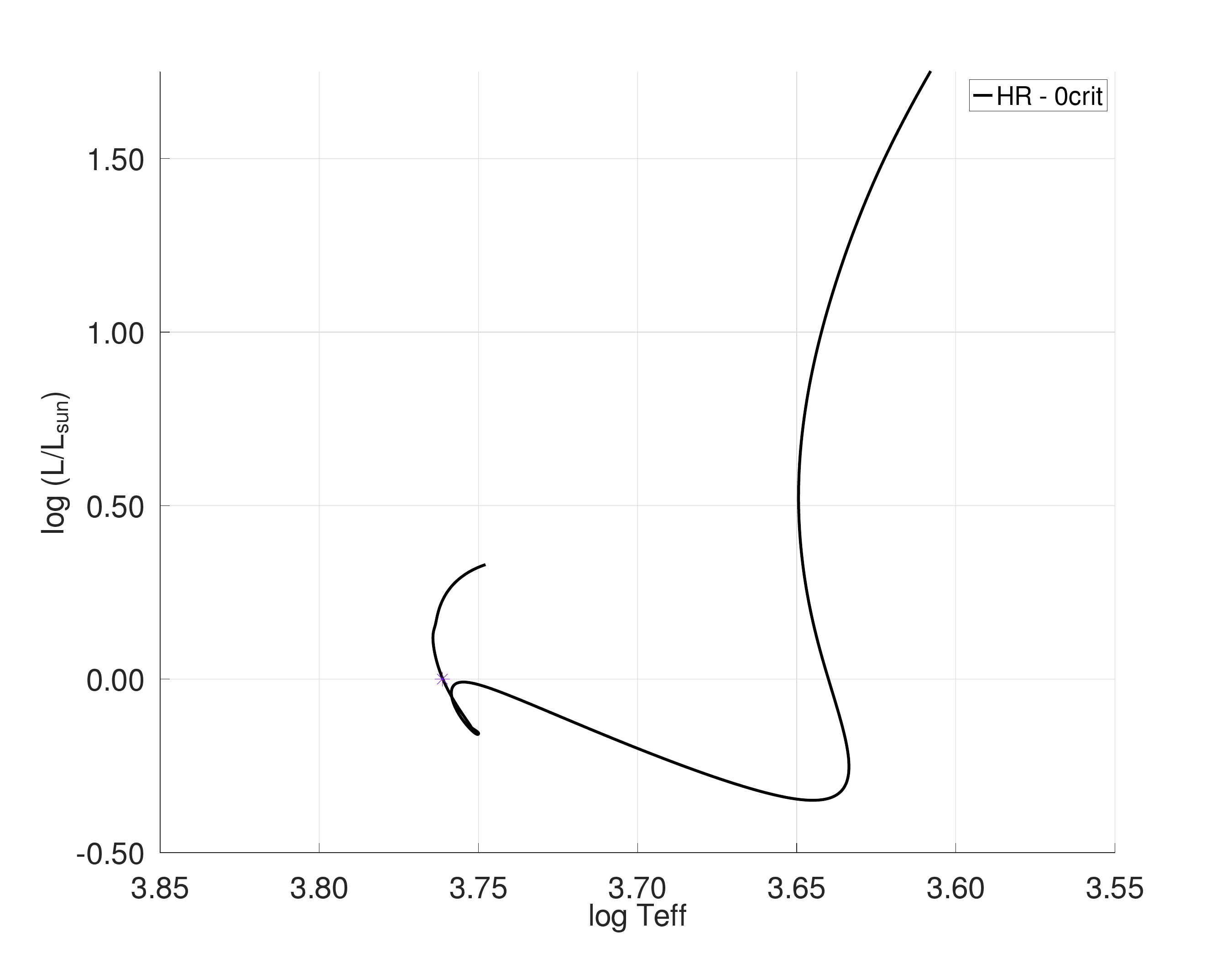}
    \caption{The calibrated solar 1$\msun$ model HR diagram. The rotation, overshooting, or magnetic braking effects are deactivated in the model. The luminosity is expressed in terms of $\lsun$.}
    \label{fig:hr_var_vel_0_0g0_cal}
\end{figure}

Our models have been calibrated to reproduce the following observables at the age of the Sun: effective temperature, luminosity, and radius. To do this, we started with a standard model without rotation, overshooting, or magnetic braking. Table \ref{tab:target_vs_mod} informs the values reproduced by the calibrated model and the error made with respect to the target values. In the HR diagram in Figure \ref{fig:hr_var_vel_0_0g0_cal}, we can see how the model reproduces the Sun's luminosity as it passes through the MS.\par

\begin{table}
	\centering
	\begin{threeparttable}
		\begin{tabular}{llll} 
			\hline
			Parameter & Sun reference & Model value & Error\\
			\hline
            Age (Ga)& 4.57 & 4.57 & <0.1\% \\
			$\teff$ (K) & 5772 & 5772.05 & <0.001\% \\
            Luminosity (log $\lsun$) & 1.0 & 3.528 10$^{-3}$ & <0.4\% \\
            Radius (log $\rsun$) & 1.0 & 2.353 10$^{-3}$ & <0.3\% \\
			\hline
		\end{tabular}
	\end{threeparttable}
	\caption{Model calibration - The table lists the parameters used during calibration (1st column), the solar values used as reference (2nd column), the values derived from the model (3rd column), and the deviations obtained (last column).} \label{tab:target_vs_mod}
\end{table}

Following \citet{Caballero2020} we restrict our simulations to solar-type stars with a mass of $1\,\msun$. The evolutionary models of stellar interiors incorporate rotation effects, account for the magnetic braking on stellar evolution, and track the rotational history and the surface abundance of $A(\isotope[7]{Li})$ for various initial angular velocities ($\omegaini$). The main differences lie in the parameterization of the new model: we have put the focus on reducing the number of free parameters involved in the simulations. A pivotal innovation lies in the introduction of a variable magnetic field strength ($B$) and a variable $\amlt$ parameter. As for the treatment of convective overshoot mixing, a departure from the previous model is evident; we deliberately deactivate this feature in our current model (in line with the \citet{Sonoi2018} parametrization) passing from two free parameters to a (not so free) one. This isolates the impact of a variable $\amlt$ on $A(\isotope[7]{Li})$ from the effects associated with convective overshooting. Notably, convective overshoot mixing is commonly addressed through ad hoc extensions of the Mixing Length Theory (MLT), introducing an additional free parameter. The Table \ref{tab:phy_mesa} enumerates the similarities and differences in the configuration of the models used in \citet{Caballero2020} and in this paper.\par

\begin{table}
\begin{threeparttable}
	\centering
	\begin{tabular}{ll} 
		\hline
		Parameter & Adopted prescriptions and values\\
		\hline
		Solar Abundance & $X_{\odot}=0.7098, Y_{\odot}=0.274$\\ & $Z_{\odot}=0.0165$\\
		Equation of State & OPAL+SCVH+MacDonald+HELM+PC\\
		Opacity & OPAL Type I for log T $\geq$ 4 \\ & Ferguson for logT $<$ 4\\
		Reaction Rates & JINA REACLIB\\
		Boundary Conditions & ATLAS12; $\tau$=100 tables + photosphere\\
		Diffusion & Track \isotope[1]{H}, \isotope[2]{He}, \isotope[7]{Li}, \isotope[7]{Be}\\
		Rotation Schema & Differential rotation at PMS \& MS\\ & Include SH\tnote{1}  , ES\tnote{2}  , GSF\tnote{3}  , SSI\tnote{4}  , DSI\tnote{5}\\
		Thermohaline & $\alpha_{{\rm th}}=666$\\
		\textbf{Convection} & $\alpha_{{\rm MLT}}$ variable depending on $\teff$\\ & \& $\gsurf$ + Ledoux\\
            Semiconvection & $\alpha_{{\rm sc}}=0.1$\\
		\textbf{Overshoot} & $f_{{\rm ov,core}}=0.0$, \& $f_{{\rm ov,sh}}=0.0$\\
		\textbf{Magnetic Field} & B(G) variable, depending on $\rho$, $\teff$ \&  $\Omega$\\
		Mass Loss & Based on \citep{Langer1998}, $\Dot{M}_{{\rm max}} = 10^{-3} \: \msun \: a^{-1}$\\
		Angular Moment Loss & $\Dot{J} = \Omega_* \; \mwind \; \ralfven^2$\\
		\hline
	\end{tabular}
    \begin{tablenotes}\footnotesize
        \item (1) Solberg-Hoiland, (2) Eddington–Sweet
        \item (3) Goldreich–Schubert–Fricke, (4) Secular Shear Instability
        \item (5) Dynamical Shear Instability
    \end{tablenotes}
    \end{threeparttable}
    \caption{Summary of adopted physics in MESA \citep[based on][]{Choi2016,Caballero2020}. Highlighted in bold are the parameters with different configuration from referenced works.}
	\label{tab:phy_mesa}

\end{table}

The models include rotation since the PMS. We computed the evolution of $1\,\msun$ stellar models at solar initial metallicity with $\omegaini = \oomegac$ varying between $0.12$ and $0.1425$. We have adopted the following theoretical approach to establish when the star is reaching the ZAMS: as both the central \isotope[1]{H} mass fraction has been reduced by $\Delta( \isotope[1]{H})= 0.0001$ from its initial value, and the star's luminosity ($L$) is almost fully ($L \geq 0.99$) generated by the nuclear reactions of \isotope[1]{H} occurring simultaneously.\par

As expressed by Eq.~\ref{eq:mb_torque}, the amount of AML depends on $\ralfven$, $\Omega$ and $\Dot{M}$. For large $\ralfven$ values, the star undergoes a significant deceleration. The value of $\ralfven$ depends on the following varying parameters: magnetic field ($B$) intensity, $\teff$ and $\Omega$. With regards to $\Dot{M}$ and as reported in Table \ref{tab:phy_mesa}, the empirical formula developed by Langer \citep{Langer1998} was used for calculating the mass loss. For a solar-type star, the $\Dot{M}$ during MS is relatively small, about  $10^{-14}\msun \, \yr$, where $1\msun \, \yr = 6.3x10^{25} \gs$ \citep{Noerdlinger2008}. \par

MESA assigns an $\Omega$ value for each cell $k$ ($\Omega_k$) which is adjusted so that the resulting angular momentum is retained after calculating the new mass of the cell $k$ ($m_k$) and its distance to the center of the star ($r_k$). After that, an AM value is assigned to each cell $k$ ($J_k$). At this point, our MB routine will turn on if the star has both an extensive convective layer and a radiative core. From this moment on, the MB routine is activated, acting as an additional mechanism to those existing in the MESA evolutionary code that participated in the star AML and modifying $J_k$. This was done by providing an additional contribution ($\Dot{J}_{k}$). This contribution is the result of the external torque exerted by the magnetic field once it has been distributed among the different layers that make up the CZ as dictated by Eq.~\ref{eq:k_jdot}:\par

\begin{ceqn}
\begin{align}
    \Dot{J}_{k} &= \Dot{J}_*\;\frac{m^{}_{k} r^2_{k}}{m^{}_* r_*^2} \label{eq:k_jdot}
\end{align}
\end{ceqn}

Finally, the effects of the AML on the Li abundances are then compared with the stars at different evolutionary states in the OCs referenced in the Table \ref{tab:oc_reduced_list}.\par

\section{Results} \label{sec_3}
The impact of rotation on both PMS and lithium depletion for solar-type stars has been widely debated in the past \citep{Pinsonneault1997,Jeffries2004,Somers2014} and revised more recently based on the availability of more precise measurements \citep{Bouvier2018, Caballero2020}. The results suggest a connection between the rotational speed of stars and the abundance of lithium detected in their stellar atmospheres, such that faster rotating stars destroy less lithium than slower rotating stars. In some ways, the results published by \cite{Eggenberger2012, Bouvier2016} are unexpected, as they point in a direction diametrically opposite to previous work, which predicted that faster rotating stars should destroy a greater amount of lithium \citep{Somers2015}.\par

To explain this pattern, it is necessary to resort to additional mechanisms linking lithium depletion and rotation. This apparent contradiction in the results of different works need not be such. In fact, different proposals are put forward that may help to understand the discrepancies, among which is the coupling of the stars with their protostellar disk \citep{Israelian2009, DelgadoMena2014, Figueira2014, Mishenina2016} that would influence the mixing mechanisms caused by the rotation \citep{Bouvier2008, Eggenberger2012}, or the influence of magnetic fields \citep{Eggenberger2009} that have the property of transmitting angular momentum (AM) much more efficiently than inducing mixing \citep{Denissenkov2007}. As a consequence of this increase in AM transport efficiency, the amount of differential rotation between the radiative and convective zones of the star is reduced (a solid-body rotation is promoted) as well as the induced rotational instabilities.\par

\begin{figure*}
	\centering
	\begin{subfigure}[h]{0.47\textwidth}
		\includegraphics[trim = 15mm 10mm 15mm 10mm, clip,width=\textwidth]{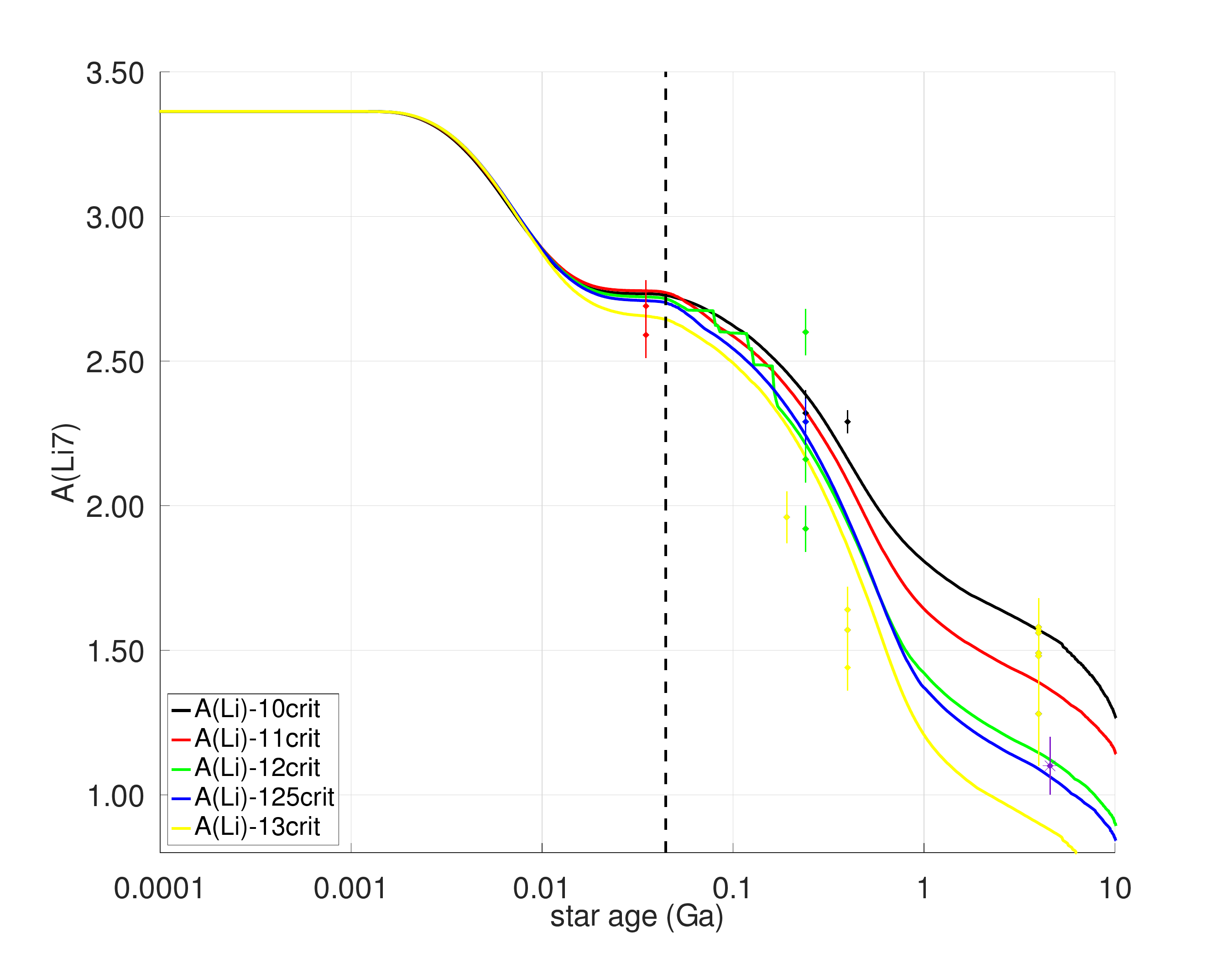}
        \caption{}
		\label{fig:li_var_vel_var_g_3}
	\end{subfigure}    
	\begin{subfigure}[h]{0.47\textwidth}
        \includegraphics[trim = 15mm 10mm 15mm 10mm, clip,width=\textwidth]{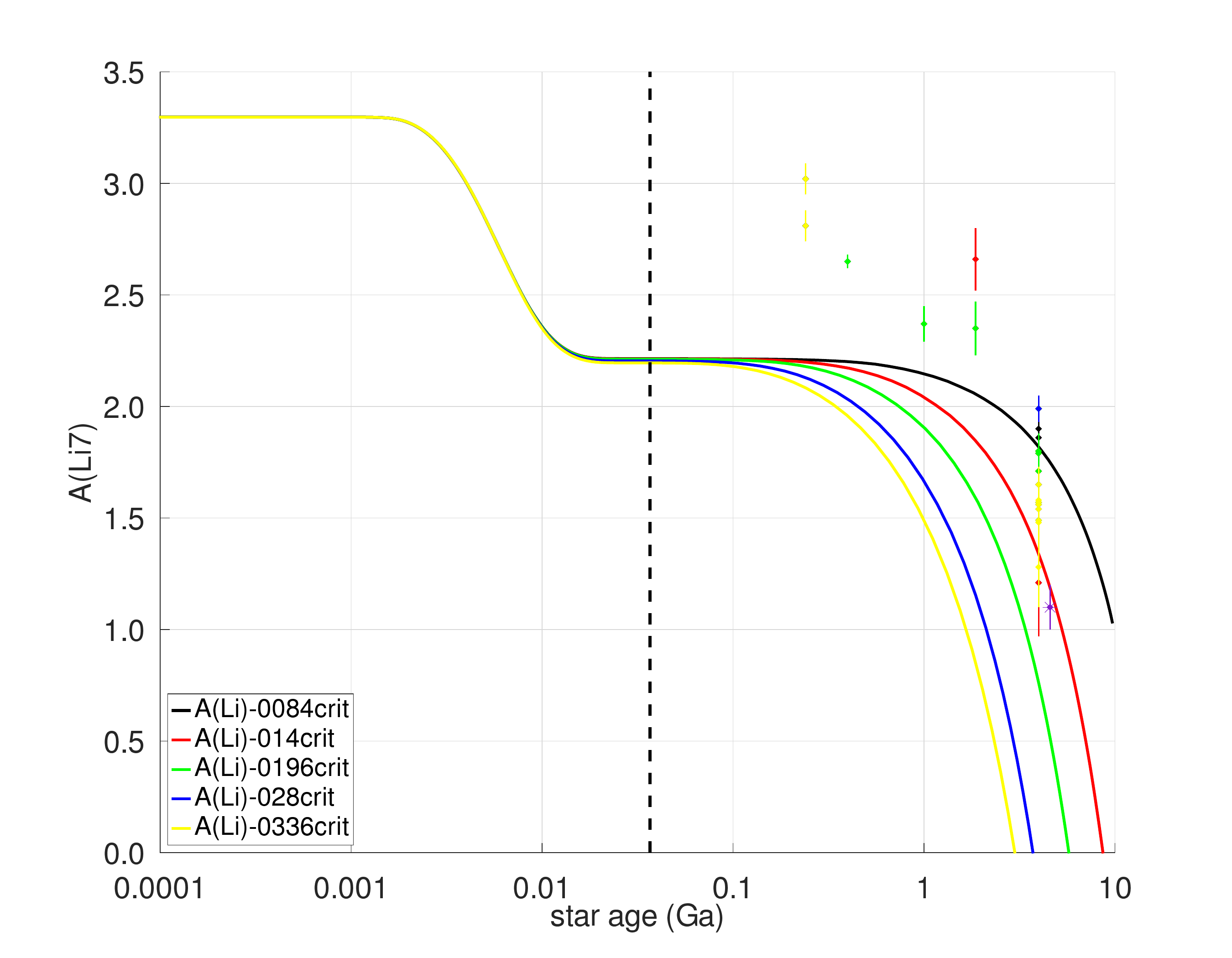}
        \caption{}
		\label{fig:li_var_vel_0g}
	\end{subfigure}
	\caption{Ratio of surface \isotope[7]{Li} abundance to \isotope[1]{H} as a function of time is depicted for various 1 $\msun$ models with magnetic field. In the figure (a), the models have initial rotation rates ranging from 0.10 to 0.13. The other colored dots represent surface \isotope[7]{Li} abundances for stars with parameters within the specified selection intervals, corresponding to the evolution curve of the same color. In the figure (b), the magnetic field were not simulated. The rest of lines are models which include initial rotation with $\omegaini$ between 0.0084 and 0.0336, respectively. The purple star is the surface Li abundances for the present-day Sun \citep{Asplund2009}. The dashed vertical line indicates makes reference to the ZAMS.}
	\label{fig:grid_caballero_2020}
\end{figure*}

\begin{figure}
	\includegraphics[clip,width=\columnwidth]{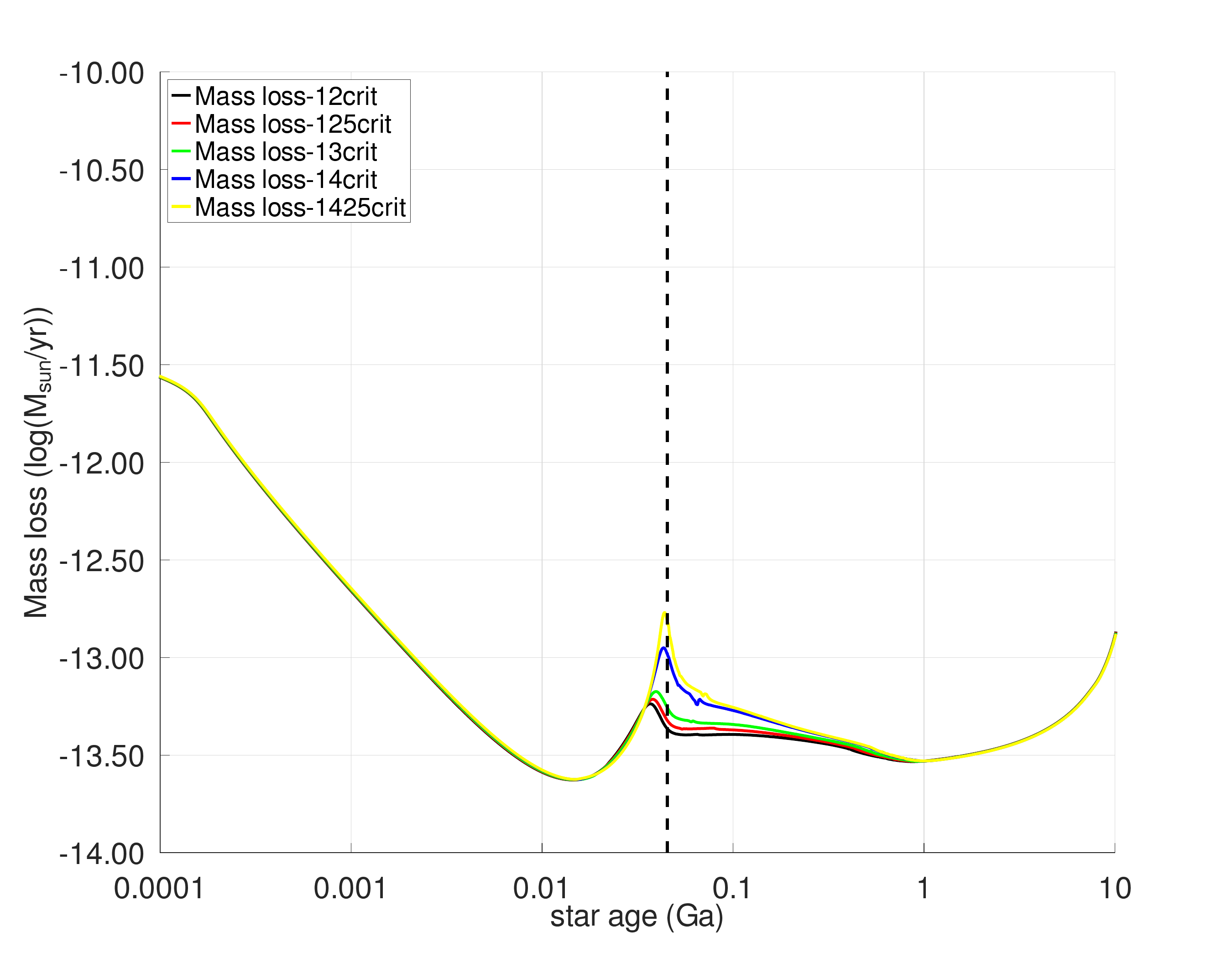}
    \caption{The evolution of mass loss $\Dot{M}$ as a function of time for several 1 $\msun$ models. The models include variable magnetic field intensity and initial rotation with $\omegaini$ between 0.12 and 0.1425. The dashed vertical line makes reference to the ZAMS.}
    \label{fig:mdot_var_vel_g3}
\end{figure}

\begin{figure}
	\includegraphics[clip,width=\columnwidth]{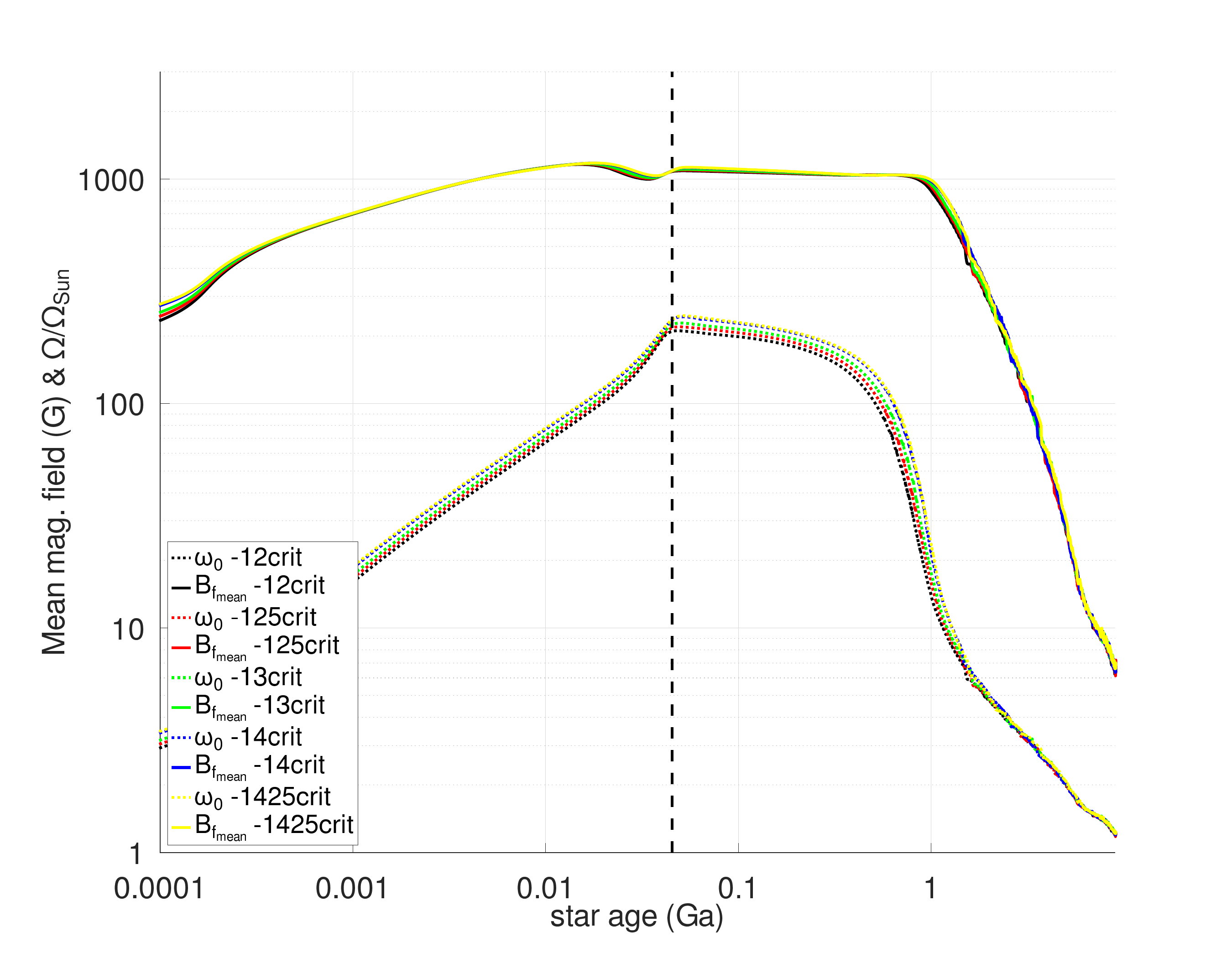}
    \caption{The evolution of magnetic field intensity and $\omegaini = \oomegac$ as a function of time for several 1 $\msun$ models. The models include initial rotation with $\omegaini$ between 0.12 and 0.1425. The solid lines represent the magnetic field strength, while the dotted lines represent the angular evolution of the star. The dashed vertical line makes reference to the ZAMS.}
    \label{fig:mag_field_var_vel_g3}
\end{figure}

In Figure \ref{fig:li_var_vel_var_g_3} it is shown the temporal evolution of surface Li abundance for several 1 $\msun$ models. Those models were initialized with different rotational velocities and took into consideration the effects of MB caused by a variable magnetic field. If we compare it with Figure \ref{fig:li_var_vel_0g} in which the effects of MB were neglected, we notice how the profiles of Li abundance were altered across PMS and MS. During the PMS we can describe the effect as modest, somewhat expected and in line with the fact that the AML caused by MB (see Eq.~\ref{eq:j_dot}) depends directly on the the $\Omega$ evolution and mass loss rate. If we take into account that for solar-type stars the models predict a modest total mass loss rate, that value is even much lower in this phase. It is in the approach phase to the ZAMS that the angular velocity reaches its maximum. This, according to our model, plays a crucial role in both the mass loss and the magnetic field strength. The higher the $\Omega$, the higher the $\Dot{M}$  (see Figure \ref{fig:mdot_var_vel_g3}), and the higher the magnetic field strength (see Figure \ref{fig:mag_field_var_vel_g3}). These effects combine and lead to an accentuation of the magnetic braking effect. As a consequence, there is a slowing down of Li destruction once the models enter the MS. The results of our simulations for those models respectively initialized to $\omegaini$ = 0.10 and 0.13 show A(\isotope[7]{Li}) compatible with that measured in the Sun surface (1.1 $\pm$ 0.1 dex). The latter gives a value of 1.12 dex, which would represent a deviation of about 3\% from the nominal value (see Figure \ref{fig:li_var_vel_var_g_3}).\par

\begin{figure}
	\includegraphics[clip,width=\columnwidth]{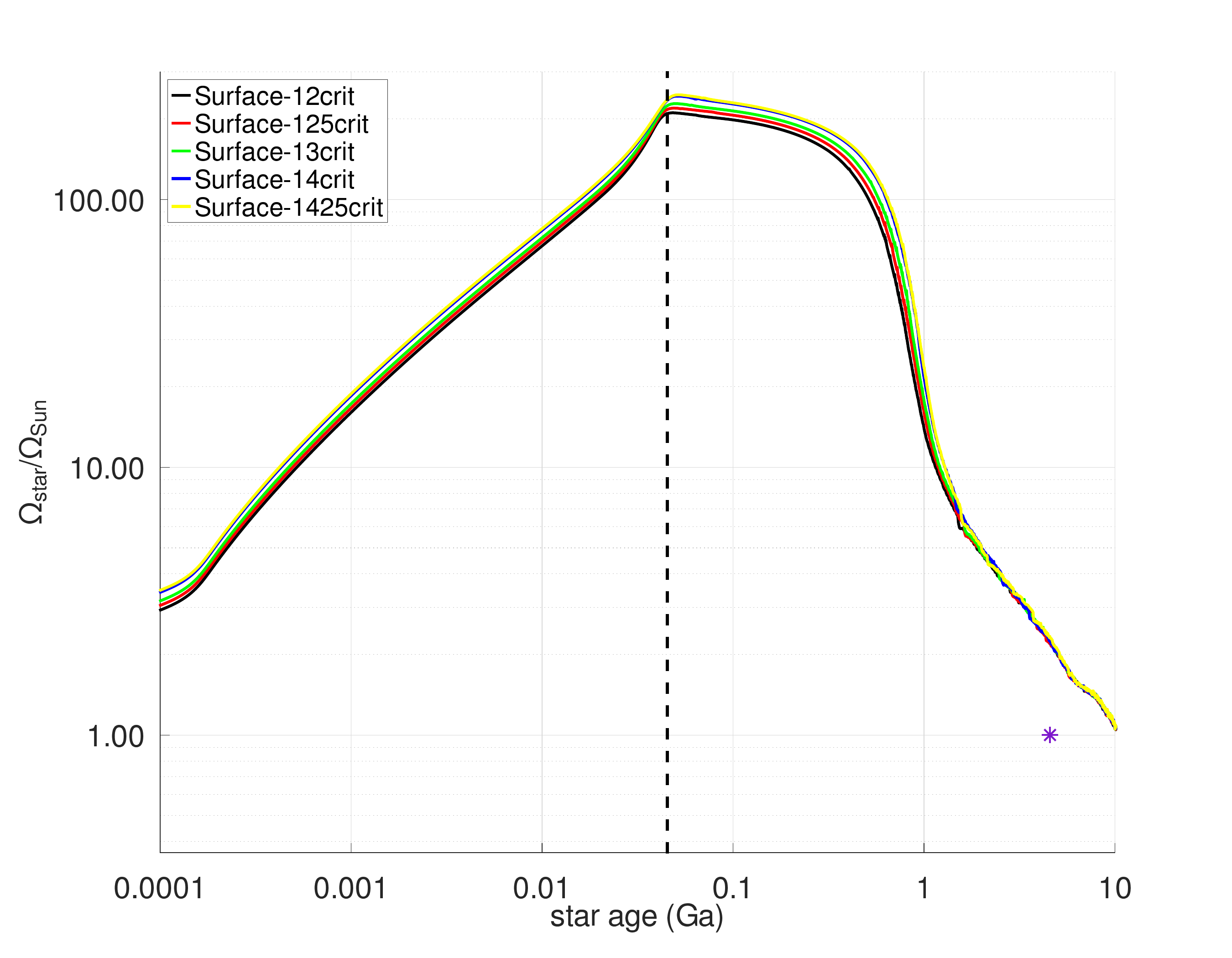}
    \caption{The evolution of surface rotational velocity, as a function of time for several 1 $\msun$ models. The models include a magnetic field with variable intensity, initial rotation with $\omegaini$ between 0.12 and 0.1425, respectively and MB. The purple star is the surface angular velocity for the present-day Sun \citep{Gill2012}. The dashed vertical line makes reference to the ZAMS.}
    \label{fig:rot_vel_var_vel_var_g3}
\end{figure}

\begin{figure}
	\includegraphics[clip,width=\columnwidth]{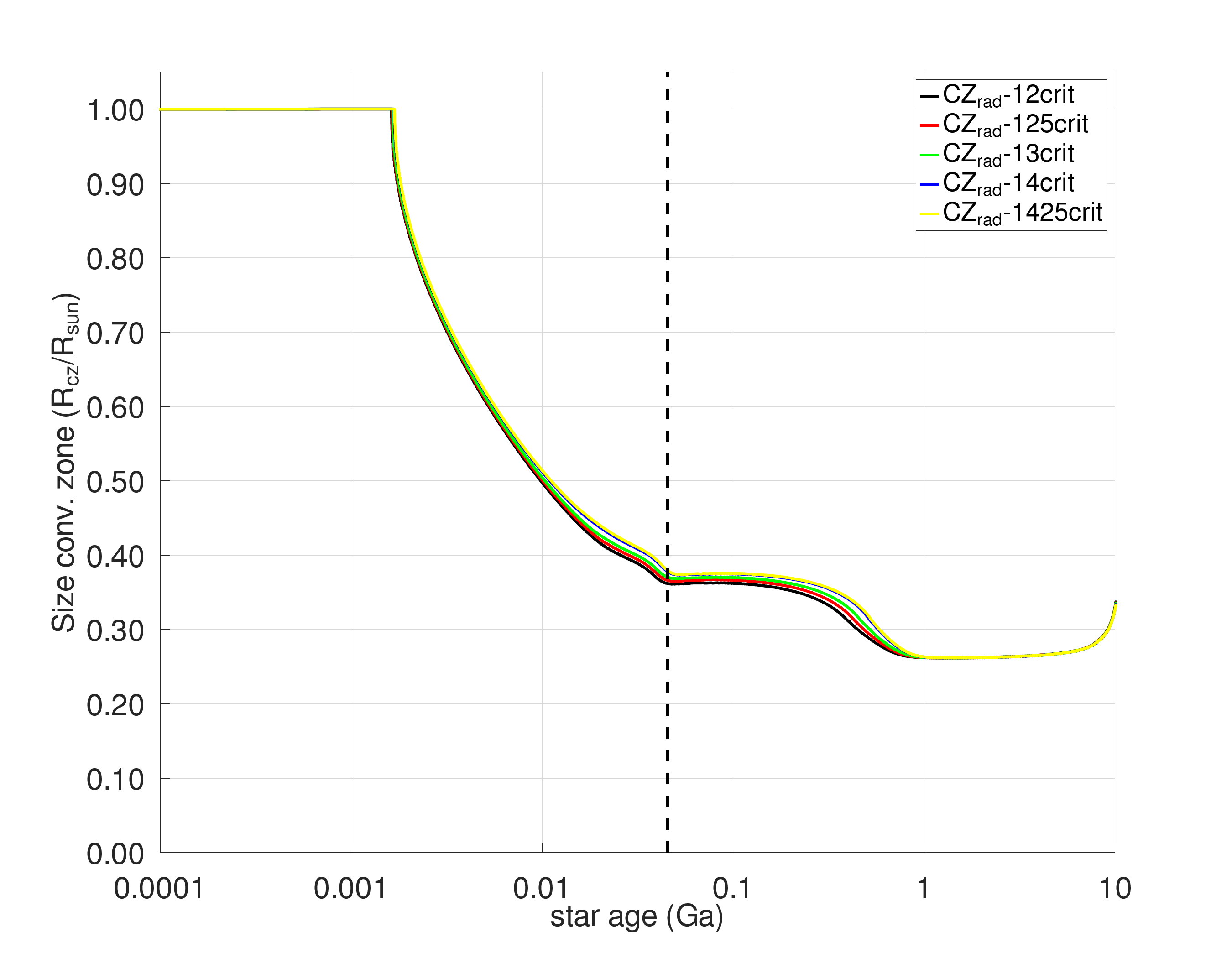}
    \caption{The evolution of the convective zone size as a function of time for several 1 $\msun$ models. The size is expressed in terms of $\rsun$. The models include a magnetic field with variable intensity, initial rotation with $\omegaini$ between 0.12 and 0.1425. The dashed vertical line makes reference to the ZAMS.}
    \label{fig:cz_var_vel_var_g3}
\end{figure}

In Figure \ref{fig:rot_vel_var_vel_var_g3} we depict the rotation profiles of stars for the surface and for the bottom of the convective envelop. It shows how the angular velocity increases in value as the star decreases in radius along the PMS. During this stage of its evolution, the star has a nearly convective structure, as can be seen in Figure \ref{fig:cz_var_vel_var_g3}. As it approaches an age of $\approx 10^6$ years, the convective zone of the star begins to shrink in size and the core of the star reaches the temperature, pressure and density conditions necessary to develop a radiative core. In turn, the difference in the angular velocities of the different models becomes more evident, being higher in those that were initialised with a higher initial angular velocity. This fact implies that the mass loss is also higher for those stars that rotate faster. As we have described above, a larger mass loss also implies a larger angular momentum loss. The combined effects of the increases in effective temperature, angular velocity and mass loss cause the magnetic field strength to reach its maximum in the approach phase of the ZAMS (see Figure \ref{fig:mag_field_var_vel_g3}). We observed that the models with lower angular velocity generally ended up exhibiting higher values for the Li abundance on the surface (see Figures \ref{fig:li_var_vel_var_g_3} \& \ref{fig:li_var_vel_var_g_1}).\par

For the models in which we obtain values of A(Li) in line with those of the Sun ($\omegaini$ = 0.14 and 0.1425), we have on the other hand measurements for the $\Omega$ and $B$ which are not are not in tune with the respective solar ones. The Sun surface rotational velocity is 2km/s with a minimum error of 3m/s, and the mean magnetic field strength is 1G. For the simulation with $\omegaini$=0.1425, we obtain a rotational velocity at equator of 4.72 km/s. Although it is a measure on the same order of magnitude, it represents a 235\% deviation. For the average magnetic field strength we have 36.9G, far away from the reference value.\par

\begin{figure}
	\includegraphics[clip,width=\columnwidth]{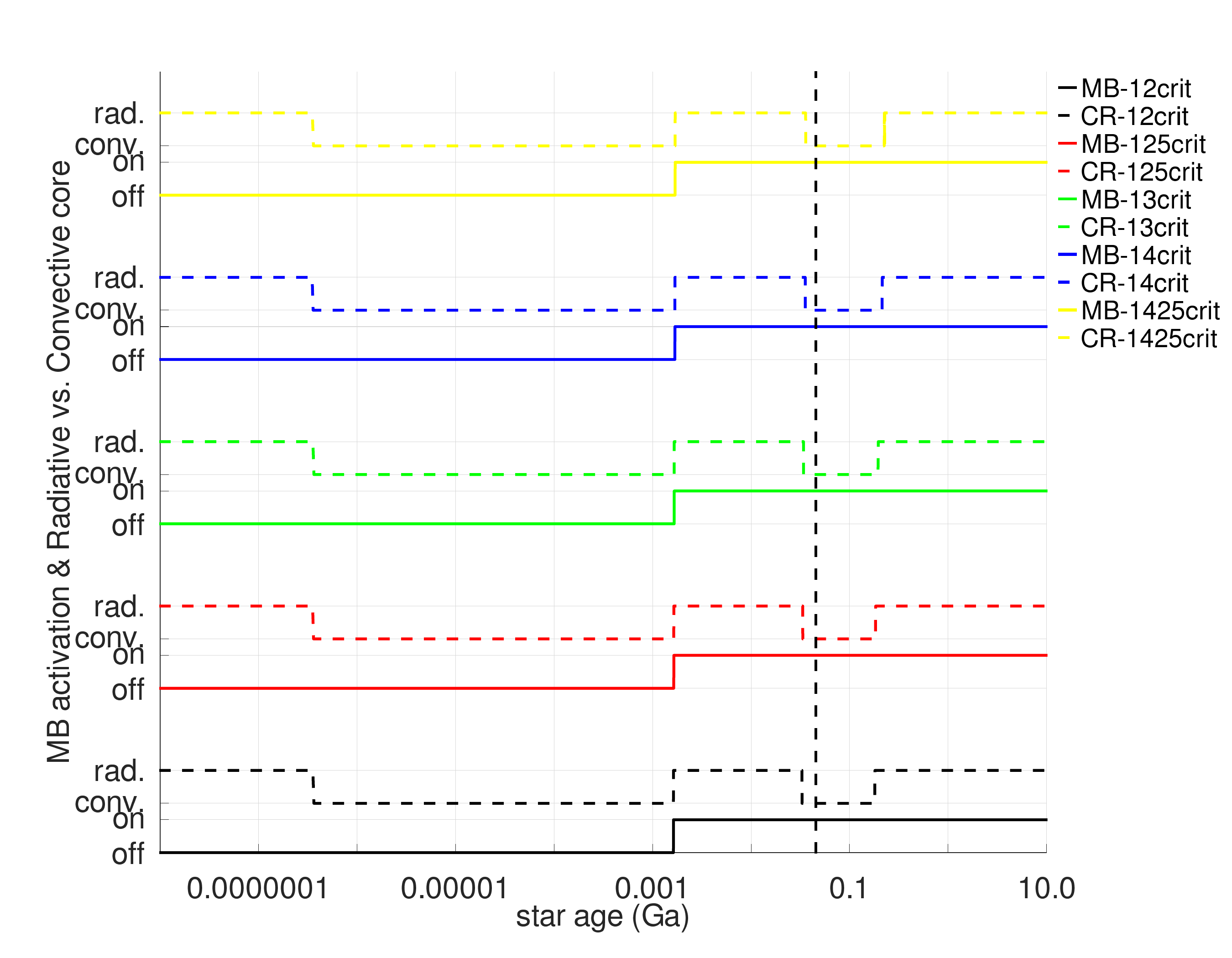}
    \caption{The activation of the magnetic braking routine as a function of the presence of a radiative core. The solid lines signal the magnetic braking routine activation (on) and deactivation (off). The horizontal dashed lines inform about the star's core nature: radiative (rad) or convective (conv). By implementation decision, once the routine is activated, it remains on even if the star's core nature change to convective. The dashed vertical line makes reference to the ZAMS.}
    \label{fig:mb_act_var_vel_g3}
\end{figure}

As described in section \ref{mod_mb}, the MB routine, which is activated when the star develops a radiative core and a convective zone above it (see Figure \ref{fig:mb_act_var_vel_g3}), distributed the total amount of AML calculated according to Eq.~\ref{eq:k_jdot} among the different layers that composed the CZ. In Figure \ref{fig:cz_var_vel_var_g3} we can observe the evolution of the most external CZ normalized with respect to the radius of the star for several 1 $\msun$ models. In accordance with the established models of stellar evolution, in a solar-type star the CZ covers practically all of it for a large part of the PMS. From the ZAMS on-wards, its size starts to decrease gradually in a first phase and then becomes more pronounced at around $4.0x10^8$ years of age. As can be seen in Figure \ref{fig:rot_vel_var_vel_var_g3}, the star, after reaching its maximum angular velocity when passing the ZAMS, starts to decelerate due to magnetic braking.\par

The CZ size is not constant as a fraction of the stellar radius; it evolves as the radiative core develops and as rotation modifies hydrostatic balance. Magnetic braking reduces rotation, lowering centrifugal support and allowing slight contraction, which accelerates CZ retreat. Changes in radius alter the $\gsurf$, $\teff$ and therefore the pressure–temperature stratification that sets the BCZ depth (see section \ref{mod_mltalpha}). As a result, the shellular approximation, which includes first-order rotational distortion in 1D models, captures these effects robustly and explains their impact on Li burning.\par

It is the presence of the MB that prevents the star from further increasing its angular velocity. In this way, the Li destruction is attenuated by a lower rotational velocity. The models start to decelerate gradually once the ZAMS is passed and during the initial period of their stay in the MS. The magnetic field strength remains close to its maximum, decreasing slightly and then decreasing more sharply. Its behaviour reflects that of the evolution of the angular velocity, a consequence of the effects of the MB. The deceleration process continues progressively until, around the present age of the Sun, we obtain angular velocity values very similar to the Sun i.e. $2.87x10^{-6}$ rad/s for the Sun vs. $6.93x10^{-6}$ rad/s for the simulation. The deceleration leads to a reduced effect of centrifugal forces, which allows for a contraction of the star's radius, and thus a smaller size of the CZ. Additionally, a lower angular velocity has an effect on the magnetic field strength, as can be seen in Figure \ref{fig:mag_field_var_vel_g3}. The coupling between angular velocity, magnetic field strength, and star radius is evident and consistent.\par

\begin{figure}
	\includegraphics[clip,width=\columnwidth]{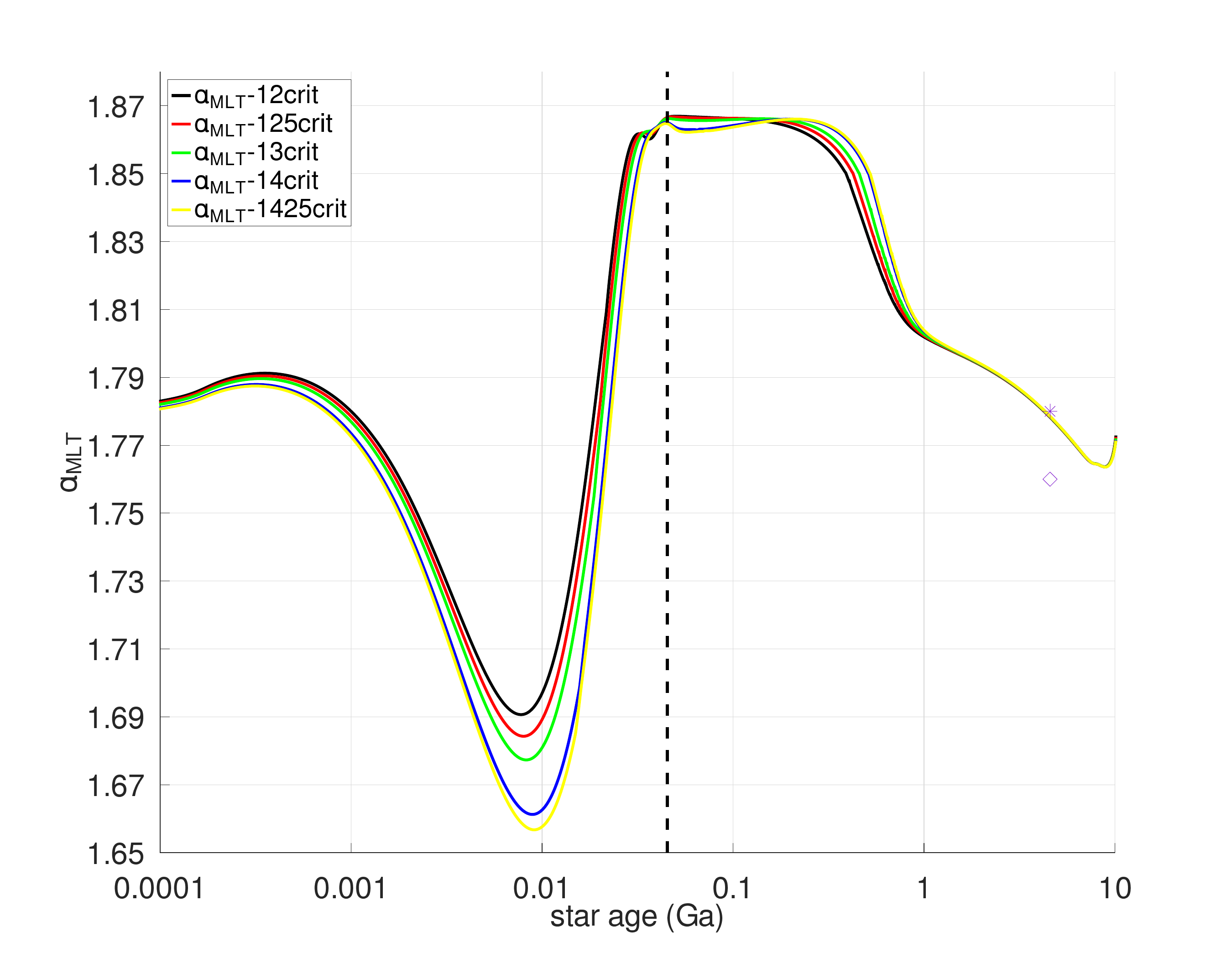}
    \caption{The evolution of $\amlt$, as a function of time and $\omegaini$ for several 1 $\msun$ models. The models include initial rotation with $\omegaini$ between 0.12 and 0.1425. The dashed vertical line makes reference to the ZAMS. The purple star and diamond are the $\amlt$ given by \citet{Sonoi2018} and \citet{Samadi2005}, respectively.}
    \label{fig:alpha_mlt_var_vel_g3}
\end{figure}

The effects of rotation and magnetic braking can also be seen in the H-R diagram and in the evolution of the $\amlt$. The inclusion of rotation implies the appearance of centrifugal forces that affect both the structure of the star, the chemical composition of the different strata that compose it, as well as its temperature and luminosity. The differential rotation between the boundaries of the radiative core and the convective layers leads to mixing effects in the so-called tachocline. The outcomes of this mixing and its hydrostatic impacts are controlled by the MLT, where $\amlt$ assumes a crucial role. As previously stated, the arbitrariness of the $\amlt$ value introduces a level of uncertainty into the MLT. Figure \ref{fig:alpha_mlt_var_vel_g3} shows how the $\amlt$ parameter evolves over time.\par 

\begin{figure*}
	\centering
	\begin{subfigure}[h]{0.47\textwidth}
		\includegraphics[trim = 15mm 10mm 15mm 10mm, clip,width=\textwidth]{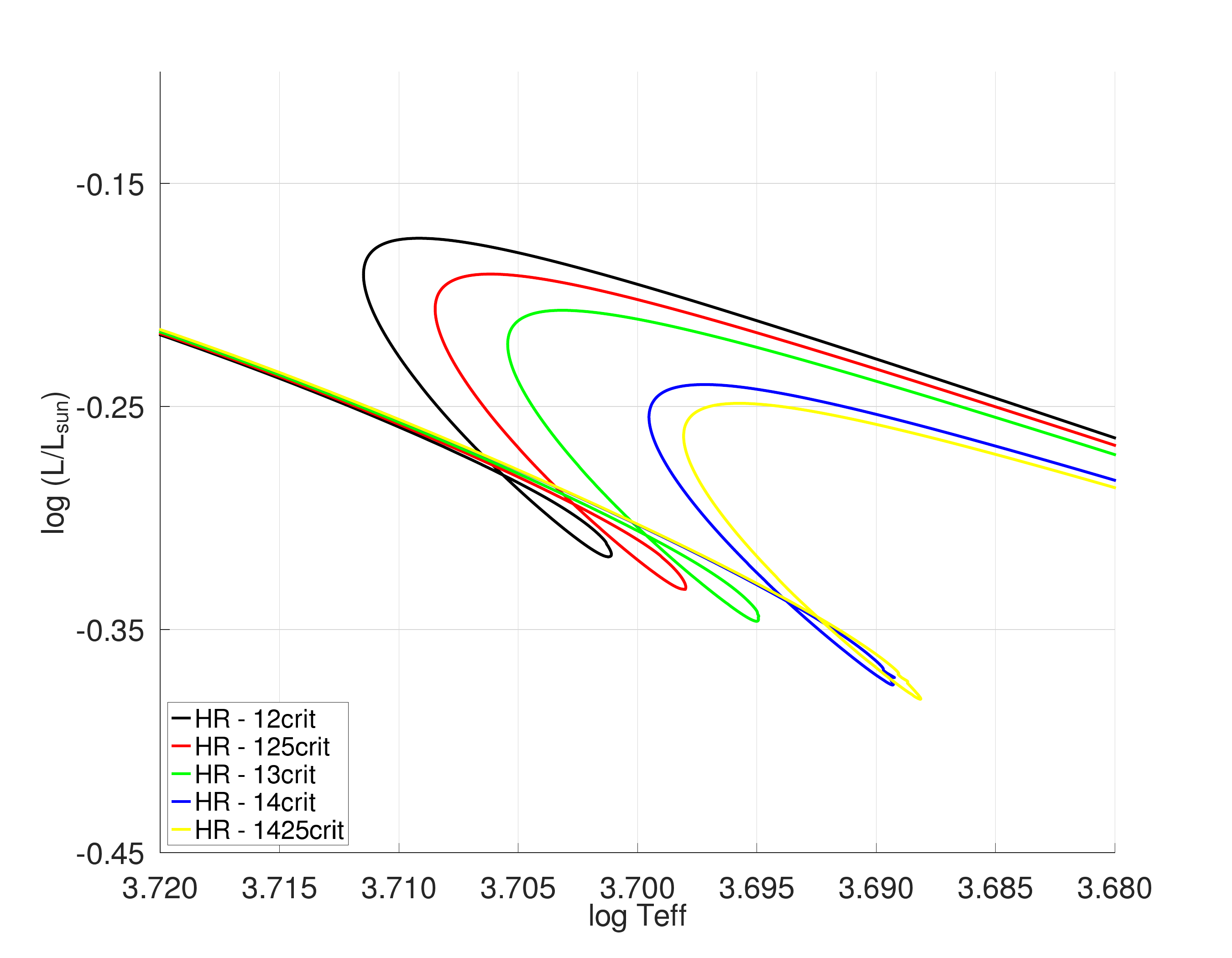}
        \caption{}
		\label{fig:hr_var_vel_var_g_z13}
	\end{subfigure}    
	\begin{subfigure}[h]{0.47\textwidth}
        \includegraphics[trim = 15mm 10mm 15mm 10mm, clip,width=\textwidth]{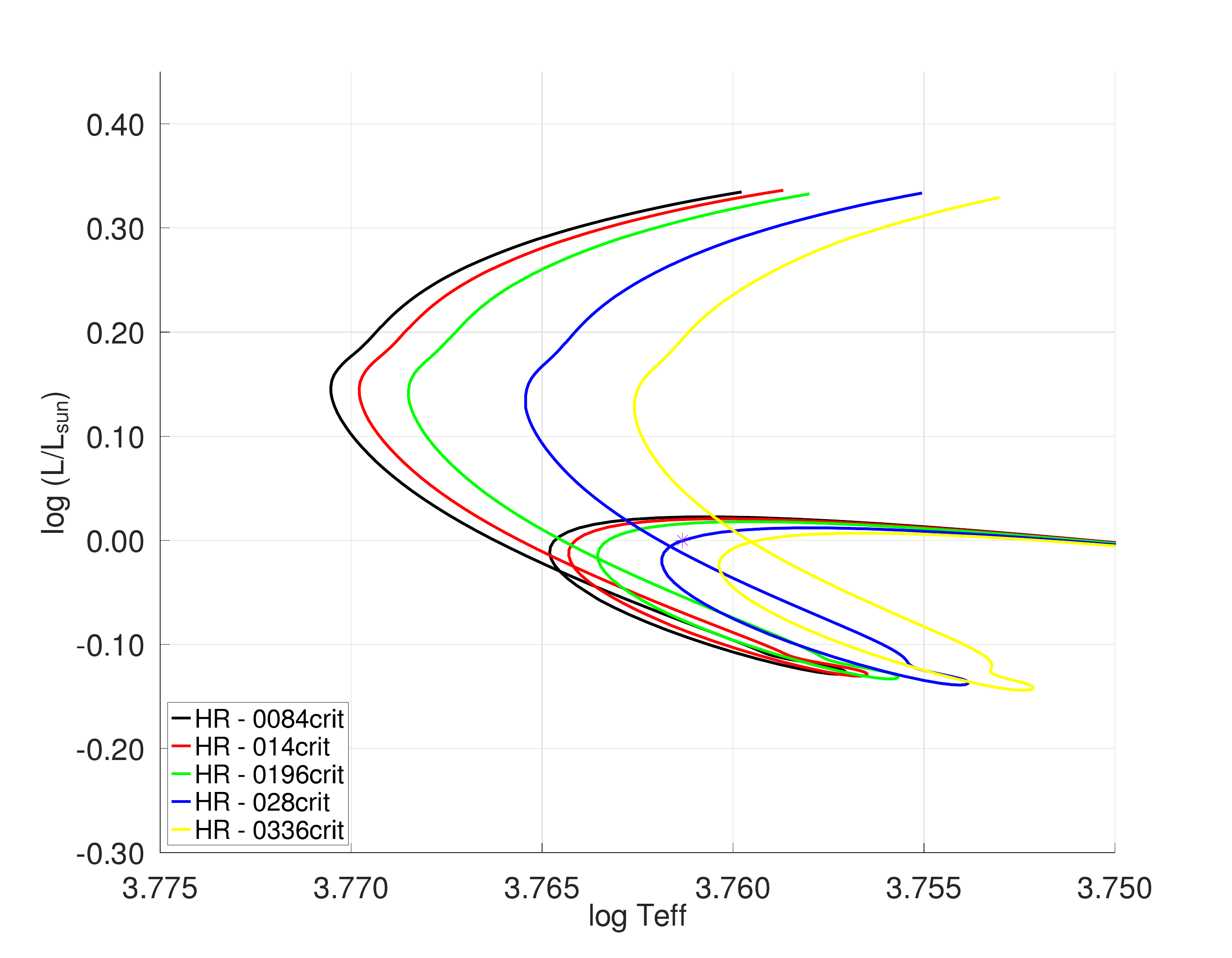}
        \caption{}
		\label{fig:hr_var_vel_0g}
	\end{subfigure}
	\caption{An example solar 1$\msun$ grid of stellar evolutionary tracks covering a range of angular velocities. The figure (a) shows in detail the combined effects of the gravity darkening and magnetic braking on the evolutionary tracks. The models include a magnetic field with variable intensity, initial rotation with $\omegaini$ between 0.12 and 0.1425. The presence of a magnetic field produces hotter stars due to the influence of the magnetic braking on the rotational velocity of the star. The figure (b) shows the stellar evolutionary tracks in the absence (0G) of a magnetic field. The rotation is activated in the models in the PMS and those models reach before the ZAMS and at a lower $\teff$ than the non-rotating one (solid black line). The luminosity is expressed in terms of $\lsun$ = 3.761. Figure adapted from \citet{Caballero2020}.}
	\label{fig:grid_hr}
\end{figure*}

Additionally, these same centrifugal forces, accentuated in stars with a higher angular velocity, reinforce the gravity darkening \citep[see e.g. ][]{Eggenberger2012,Paxton2019,Gossage2021} as well. The centrifugal force causes the star to lose its spherical symmetry and become spheroidal, with a larger radius in the equator than in the pole. The consequence is that the pressure ($\gsurf$) to which the gas is subjected is lower when the star is spheroidal than spherical. This effect causes stars that rotate fast enough to appear less dense ($\rho$), less luminous ($L$) and therefore with a lower effective temperature ($\teff$) as they approach the ZAMS (see Figure \ref{fig:hr_var_vel_var_g_z13}). If we compare the non-rotating model (black solid line in Figure \ref{fig:hr_var_vel_0g}) with the rotating ones, we can recognize that at the end of the PMS, the latter reach the ZAMS with a lower $\teff$ than the former. This also has an impact on the evolution of $\amlt$ because, as discussed above, in our model its value depends proportionally on $\teff$ and $\gsurf$. Both parameters are comparatively smaller for stars that rotate faster than for those that rotate more slowly. As a consequence $\amlt$ yields a lower value in its time evolution in those phases in which the star rotates faster, being this effect more accentuated in the ZAMS approach.\par

\begin{figure}
	\includegraphics[clip,width=\columnwidth]{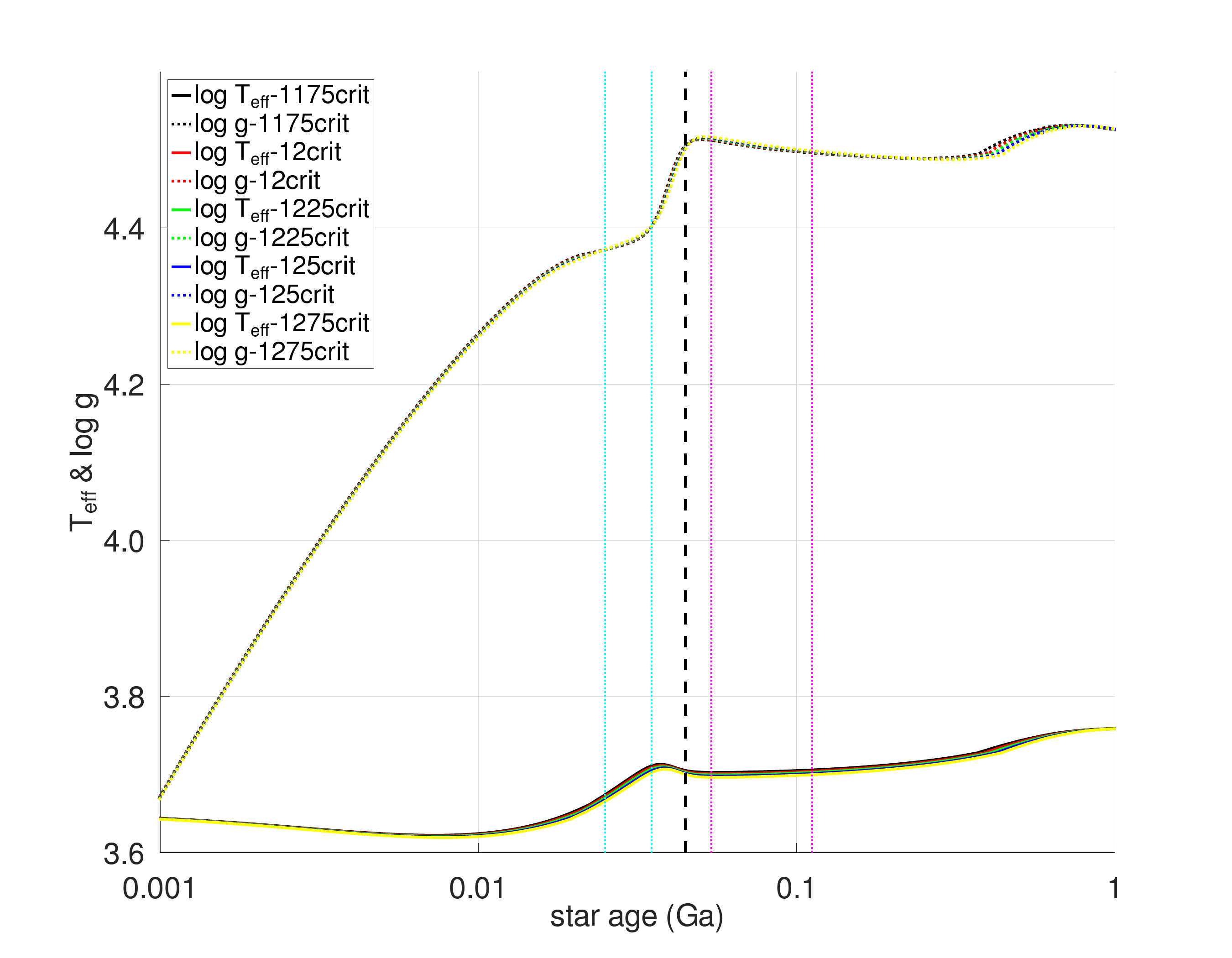}
    \caption{The evolution of $\teff$ and $\gsurf$, as a function of time and $\omegaini$ for several 1 $\msun$ models. The models include initial rotation with $\omegaini$ between 0.12 and 0.1425. The age intervals [$2.5x10^{7},3.5x10^{7}$] Ga and [$5.4x10^{7},11.2x10^{7}$] Ga, delimited by the cyan and magenta lines respectively, highlight periods in which the fastest model with $\omegaini$=0.1425 exposes a surface gravity higher than that of the rest of the slower ones. The dashed vertical line makes reference to the ZAMS.}
    \label{fig:teff_logg_var_vel_g_z13}
\end{figure}

\begin{figure}
	\includegraphics[clip,width=\columnwidth]{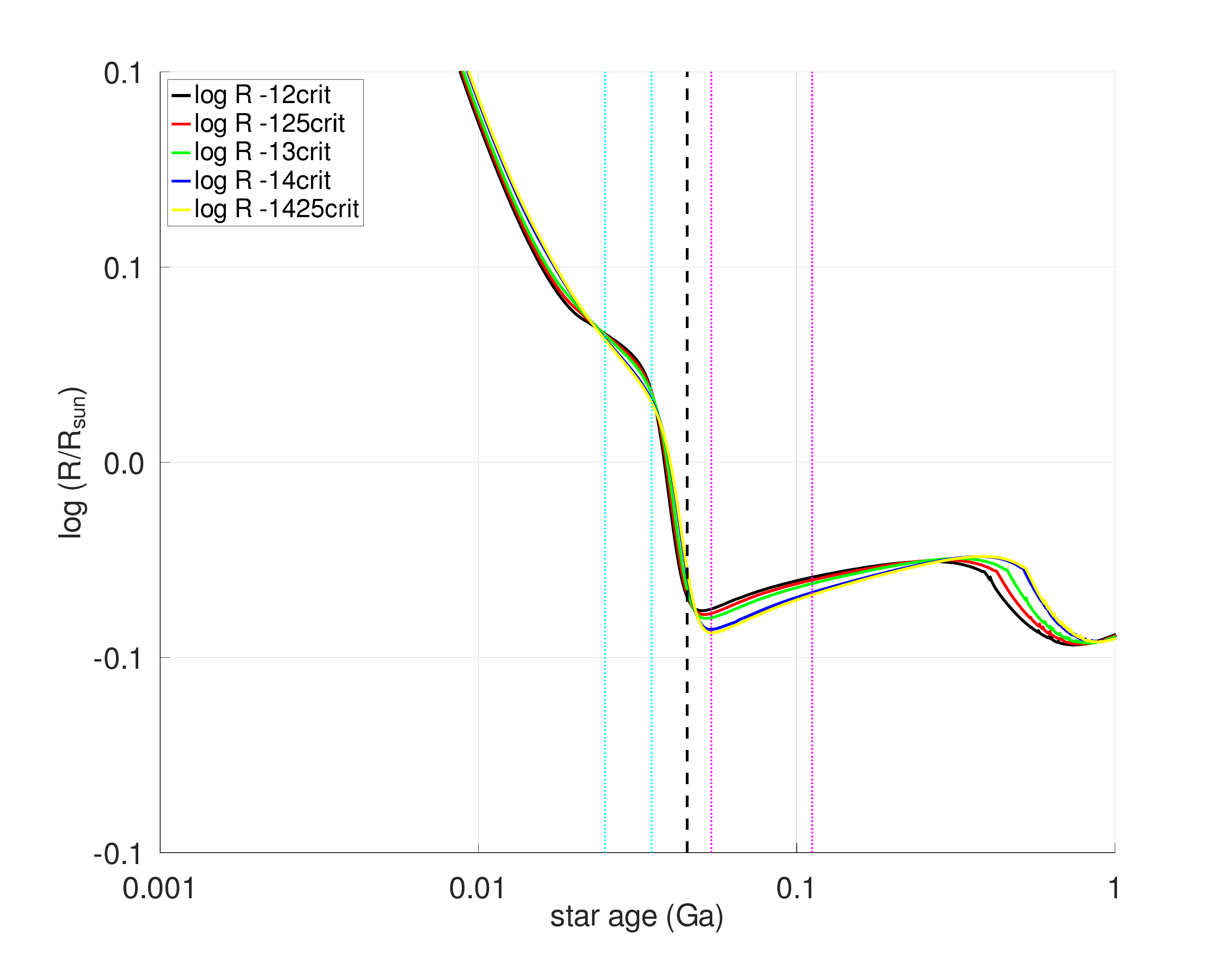}
    \caption{The evolution of the stellar radius, as a function of time and $\omegaini$ for several 1 $\msun$ models and their. The models include initial rotation with $\omegaini$ between 0.12 and 0.1425. The dashed vertical line makes reference to the ZAMS.}
    \label{fig:lograd_var_vel_g_z13}
\end{figure}

Figure \ref{fig:teff_logg_var_vel_g_z13} shows how $\teff$ and $\gsurf$ behave over the evolution of the models. For those with a higher rotation speed, we observe that both their temperature and surface gravity are lower than in the slower rotating models, in line with the gravity darkening. It is worth noting that in the ZAMS approach phase we observe for the fastest rotating model (yellow) that its surface gravity is higher than that of the rest of the slower models. This can be explained by looking at Figure \ref{fig:lograd_var_vel_g_z13}, which shows the evolution of the stellar radius for the different models, as a function of time and $\omegaini$. Around the interval $2.5x10^{7}$ and $3.5x10^{7}$ Ga (delimited by the cyan lines) and after the ZAMS, from around $5.4x10^{7}$ to $11.2x10^{7}$ Ga (delimited by the magenta lines) the stellar radius of the fastest model is smaller than that of the rest, producing a higher $\gsurf$, and this in turn means less mass loss (see Fig. \ref{fig:mdot_var_vel_g3}). Let us recall that the stellar radius has an inversely quadratic influence on the value of $\gsurf$. This "anomaly" disappears as soon as the stellar radius becomes again bigger for the fastest model.\par

\subsection{Caballero et al. (2020) comparison}
In \citet{Caballero2020}, we addressed the study of the A(Li) in the Sun. In that work, we fixed the values of $\amlt$ and $B$. In the present work, we tackle the possibility of avoiding fixed terms and making both $\amlt$ and $B$ variables with time, as expected from the stellar evolution. A comparison between both works allows us to highlight the advantages of this new approach. Table \ref{tab:caballero2024_2020} summarises the main differences between the two works.

\begin{table}
	\centering
    \begin{threeparttable}
	\begin{tabular}{lll} 
		\hline
		Parameter & This work & \citet{Caballero2020}\\
		\hline
		Convection & MLT variable $\alpha_{{\rm MLT}}$ & MLT fixed $\alpha_{{\rm MLT}}=1.82$\\
            Overshoot & deactivated & $f_{{\rm ov,core}}=0.0160$,\\ & & $f_{{\rm ov,sh}}=0.0174$\\
		Magnetic Braking & \citet{Gallet2013} & \citet{Ud-Doula2008} \\ Formalism & & \\
		Magnetic Field & B(G) variable & B(G) pre-set [3.0G-5.0G]\\ Strength & & \\
            Number of OC's & 64 & 1 \\
            Source of data & GDR3.0 GES DR5.0 & \citet{Sestito2005} \\
		\hline
	\end{tabular}
    \end{threeparttable}
    \caption{Comparison with \citet{Caballero2020} work.} \label{tab:caballero2024_2020}
\end{table}

\begin{figure}
	\includegraphics[clip, width=\columnwidth]{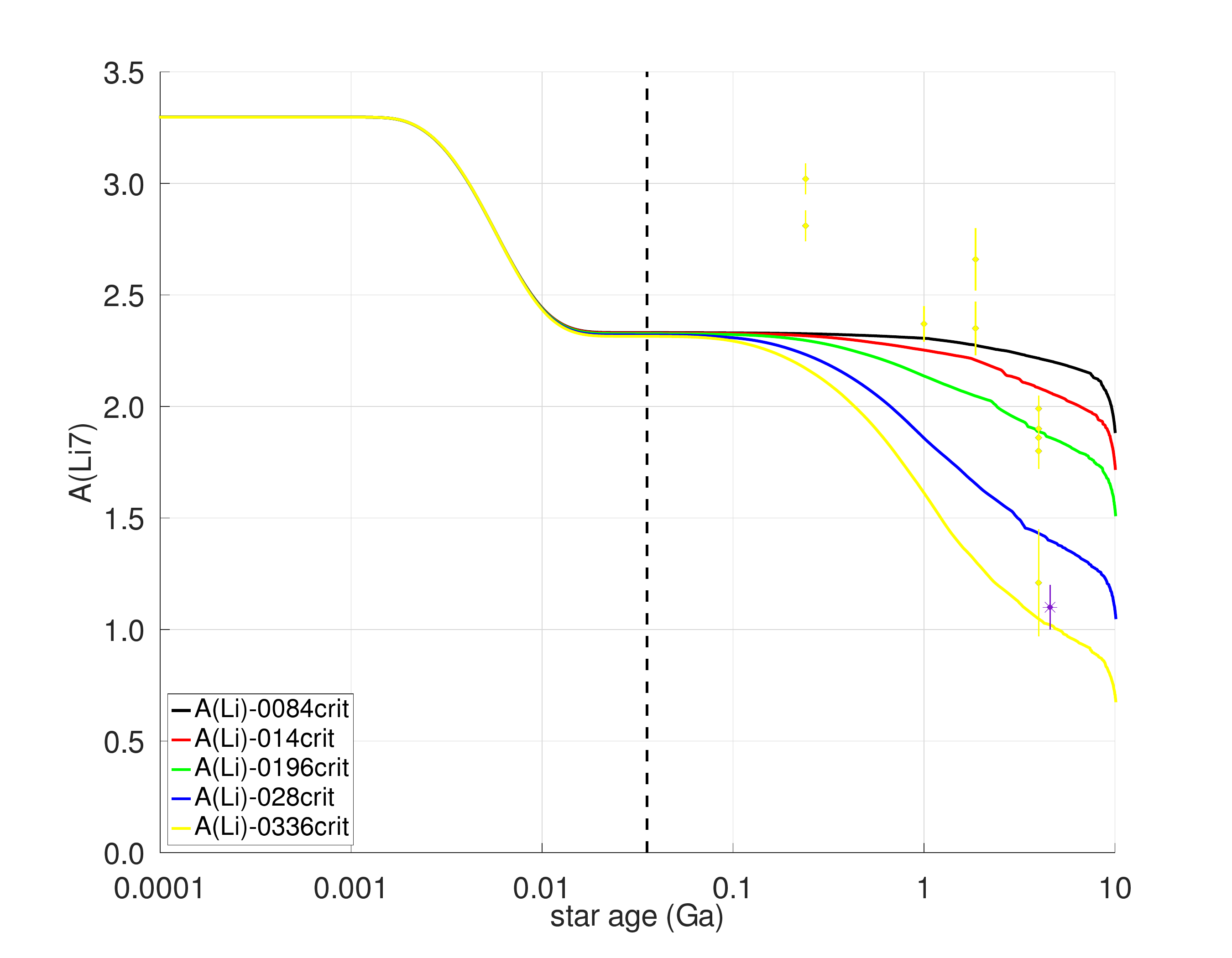}
    \caption{The evolution of surface \isotope[7]{Li} abundance relative to \isotope[1]{H}, as a function of time for several 1 $\msun$ models. The models include a magnetic field with an intensity of 4G. The solid black line represents the reference model according to \citet{Choi2016}. The rest of lines are models which include initial rotation with $\omegaini$ between 0.0084 and 0.0336, respectively. The purple star is the surface Li abundances for the present-day Sun \citep{Asplund2009}. The dashed vertical line makes reference to the ZAMS.}
    \label{fig:li_var_vel_4_0g4}
\end{figure}

\begin{figure}
	\includegraphics[clip,width=\columnwidth]{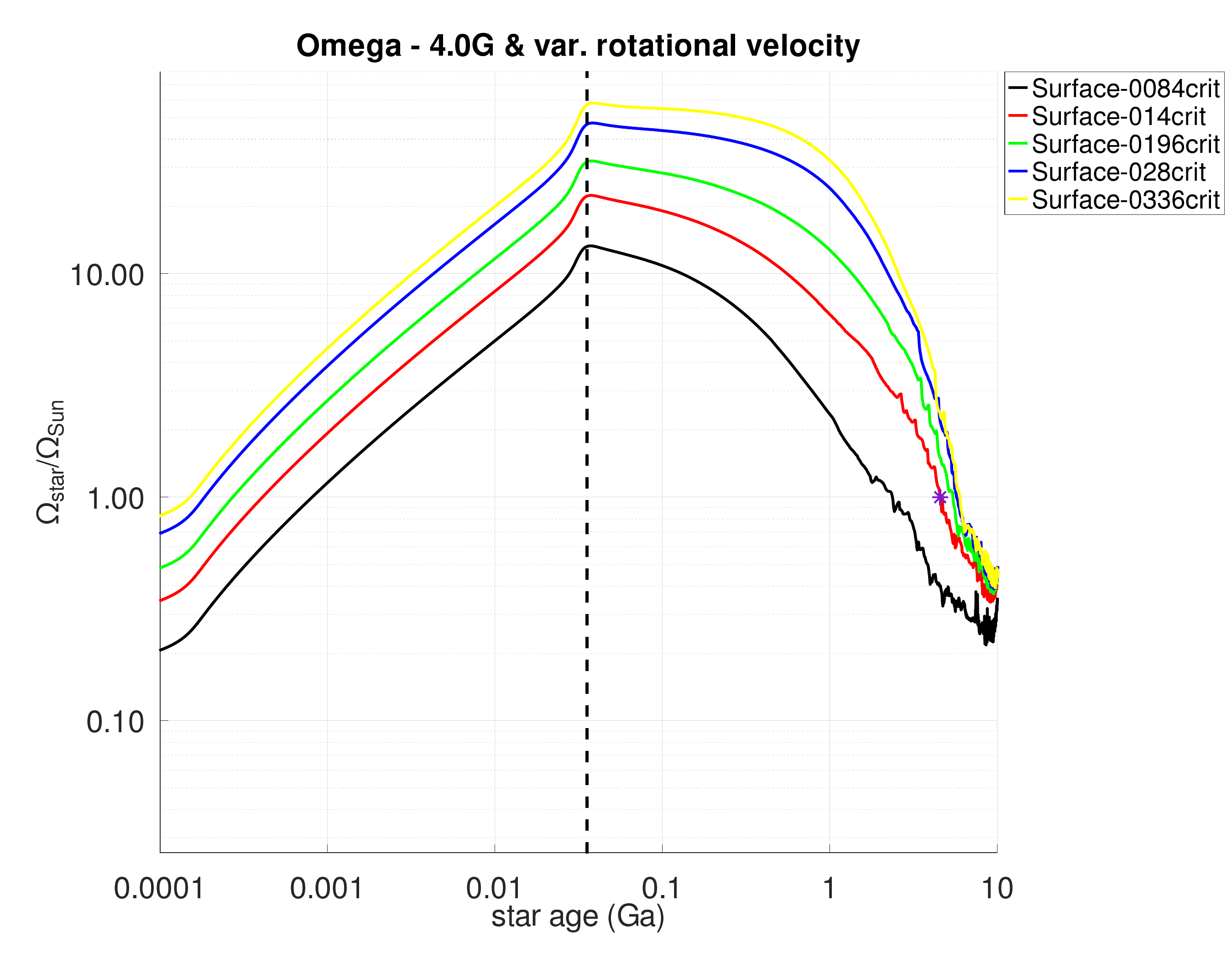}
    \caption{The evolution of surface rotational velocity, as a function of time for several 1 $\msun$ models. The models include a magnetic field with an intensity of 4G, initial rotation with $\omegaini$ between 0.0084 and 0.0336, respectively and MB. The purple star is the surface angular velocity for the present-day Sun \citep{Gill2012}. The dashed vertical line makes reference to the ZAMS.}
    \label{fig:rot_vel_var_vel_4_0g4}
\end{figure}

The Figure \ref{fig:li_var_vel_4_0g4} shows the evolution of Li abundance for $\omegaini$ between 0.0084 and 0.0336 in the presence of a constant magnetic field of $B=4$G. We have augmented the original version with the same set of OCs data gathered for the present work, so that a consistent comparison between both formalisms is possible. If we focus our attention on the simulation with $\omegaini$ = 0.0336, it reproduces  A(Li) = 1.024 dex, a value which is within the uncertainties measured for the Sun (1.1 $\pm$ 0.1 dex) and the values obtained in Figure \ref{fig:li_var_vel_var_g_3} for the present work (1.12 dex) for $\omegaini$ = 0.12 and 0.125. Although in the latter case the models were initialized at higher $\omegaini$ values than in the former, resulting in $B$ up to 3 orders of magnitude more intensive ($\approx$ 1KG versus 4G) the simulation in both works ends up producing very similar values for both the rotational velocity and the A(Li). This is partially explained because a higher magnetic field strength implies an intensive MB effect (refer to Figures \ref{fig:rot_vel_var_vel_var_g3} \& \ref{fig:rot_vel_var_vel_4_0g4}). In the former case, the deceleration effect is sharper than in the latter, so it ceases to destroy Li before.\par

However, this effect is not sufficient to effectively counteract the destruction of Li. The other decisive element that comes into play is the value of $\amlt$. As exemplified in \citet{Caballero2020} its value has an inversely proportional impact on Li destruction, where a lower value resulted in less Li destruction, particularly during PMS. In the figure \ref{fig:alpha_mlt_var_vel_g3} we observed the temporal $\amlt$ value evolution. Until the latest phase of the PMS, the value $\amlt$ is substantially less than 1.82, the pre-set and invariant value used in our previous work, resulting in less Li destruction and allowing the model to reach the ZAMS with a $\approx$ 29\% dex higher Li reservoir. This major Li reserve compensates for the greater destruction induced by the models initialized with a $\omegaini$ higher angular velocity.\par

Similarly, we show the evolution of the surface rotational velocity in Figure \ref{fig:rot_vel_var_vel_4_0g4} for the case of constant B. Again, we obtain similar values for the final rotation than in the case of variable B (see Fig.~\ref{fig:rot_vel_var_vel_var_g3}, which are explained by the higher efficiency of the MB in the latter. Although the simulations with variable B reach the ZAMS with rotational velocities up to 200\% for the models initialised with higher $\omegaini$, the stronger magnetic field produces a more pronounced deceleration for the same time interval. This fact is especially noticeable in the time period between the ZAMS and 1 Ga. The rotational velocity obtained at the equator of 4.82 km/s in the case of variable B is quite close to 4.72 km/s, that is, the value obtained for $\omegaini$ = 0.0336 in the case of constant B.\par

\subsection{Gossage et al. (2021) Comparison}
In \citet{Gossage2021}, the authors follow a similar approach to the present work. They used a couple of approaches to model the magnetic braking with the aim of reproducing observed rotational velocities and A(Li) in OCs. A difference is that our models incorporate a variable $B$ and $\amlt$, although only one approach to model the magnetic braking. They also approach the problem of AML from a similar starting point to ours, but with some notable variations. In the Table \ref{tab:gassage_vs_navarro} we compare the similarities and differences between the two studies.\par

In both works the models consider rotation not only in the MS but also in the PMS, with the distinction that in \citeauthor{Gossage2021}'s work they take into account disk-locking. This aspect has not been covered in our simulations and has been left for future work. Similarly, the effects of magnetic braking on angular momentum loss are also considered, in our case for stars with 1$\msun$ (solar twins) and in theirs with 0.1$\msun$$\le$$\mstar$$\le$1.3$\msun$.\par


These two works complement each other as they are based on different formalisms. On the one hand, our work is based on the proposal of \citet{Gallet2013} whose most relevant parameters are $\rossby$ and the presence of a magnetic field of varying strength (B), whereas Gossage presents two formalisms: \citet{Matt2015} and \citet{Garraffo2018}. The former is mainly characterised by $\rossby$ and the $\mstar$, while the latter by $\rossby$ and a parameter that accounts for the complexity of the magnetic fields, being equal to 1 for a dipole field and larger for higher-order multipoles. As far as OCs observations are concerned, here we covered a wider range of ages [0.001-6.761] Ga. Finally, both papers simulate their respective models in MESA but use different versions.\par

The most relevant difference between both works is the non-simulation of the magnetic field strength ($B$) by \cite{Gossage2021}. Although the formalism of \cite{Garraffo2018} does contemplate it, in \cite{Gossage2021} the authors opted for a simplification, which allowed not having to resort to the use of this parameter. This means that we cannot compare the results obtained in our research with theirs.\par

\citet{Gossage2021} manage to reproduce more accurately the Sun's rotation profile at its current age \citep[see][panel (a) in Figures 2, 4 \& 6]{Gossage2021} although, as the authors of the paper describe, they calibrated their models to do so, by including an additional diffusion-free parameter ($v_0$). In our work, our model does not include this type of additional parameterization and reproduces an equatorial rotation speed of 4.72 km/s, which is higher than the solar value, v = 2.0 km/s. As far as A(Li) is concerned, their model does not reproduce the value for Sun-like stars, including the Sun itself \citep[see][panel (d) in Figures 2 \& 4, and (b) in Figure 6]{Gossage2021}. By recalibrating some of the parameters of their model and/or incorporating additional mechanisms that contribute to Li destruction, the authors hope to obtain values in line with observational data. In our case (see Section \ref{sec_conclusions}), we were able to reproduce A(Li) with greater precision.

\begin{table}
	\centering
    \begin{threeparttable}
	\begin{tabular}{lll} 
		\hline
		Parameter & This work & \citet{Gossage2021}\\
		\hline
            
		Rotation & At PMS \& MS & At PMS \& MS\\
  		Solar Mass & $\mstar$=1$\msun$ & 0.1$\msun$$\le$$\mstar$$\le$1.3$\msun$ \\
		Convection & MLT variable $\alpha_{{\rm MLT}}$ & MLT fixed $\alpha_{{\rm MLT}}$\\
		Magnetic Braking & \citet{Gallet2013} & \citet{Matt2015}, \\ Formalism & & \citet{Garraffo2018}\\
		Magnetic Field & B(G) variable & Not simulated\\ Strength & & \\
            Disk locking & Not present & Present\\
            Number of OC's & 64 & 11 \\
            Source of data & GDR3.0 GES DR5.0 & Several works \\
            MESA version & r10398 & r11701\\
		\hline
	\end{tabular}
    \end{threeparttable}
	\caption{Comparison with \citet{Gossage2021} work.}\label{tab:gassage_vs_navarro}    
\end{table}

\section{Conclusions and future works} \label{sec_conclusions}
We advanced the modeling of magnetic braking and lithium (Li) depletion in solar-like stars by introducing a time-dependent magnetic field and an adaptive convective mixing efficiency parameter. As a star evolves, changes in its rotational velocity—driven by variations in radius and angular momentum—necessarily influence the stellar magnetic field. Thus, assuming a constant magnetic field throughout stellar evolution is an unrealistic simplification. Similarly, the extent and efficiency of internal convective zones evolve significantly, even appearing or disappearing over time, making a fixed mixing efficiency parameter inadequate for capturing the true mixing dynamics. These considerations critically impact the predicted Li abundances in models of the Sun and solar-like stars.\par

We adopted the variable magnetic field following the prescription of \citet{Gallet2013}, an adaptive mixing efficiency parameter based on the 3D atmosphere parametrization of \citet{Sonoi2018}, and the magnetic braking formalism of \citet{Caballero2020}. Our model predictions for Li abundance and rotational velocities were then compared with observational data from the present-day Sun and 64 open clusters, which sample the Sun’s past and future evolution.\par

Our models predict a Li abundance of 1.12 dex for Sun-like stars, which lies within the observational uncertainty of the present-day Sun (1.1 $\pm$ 0.1 dex). However, the models yield a higher equatorial rotational velocity (4.72 km/s) compared to the solar value (2.0 km/s), suggesting that additional angular momentum loss mechanisms may be needed to reconcile theory with observations. The discrepancy in rotation could also contribute to the slightly higher Li depletion in our models. Furthermore, while the models produce strong magnetic fields ($\sim$40 G) and efficient braking, the predicted field strength exceeds the Sun’s observed average ($\sim$1 G), indicating potential oversimplifications in the magnetic treatment or missing physics in the magnetic field saturation or dissipation.\par

We further examined the evolution of convective zones and. As expected from the 3D parametrization, converged to the solar-calibrated value ($\sim$1.8) by the present age, though it varied between 1.67 and 1.86 during earlier phases. While this variability influenced Li depletion at different initial rotations, the models could not fully reproduce the Li trends across all open clusters. This discrepancy likely arises from unaccounted factors, such as cluster-specific formation scenarios (e.g., binary interactions, mass transfer, planetary engulfment) or variations in initial chemical composition.\par

Given that faster rotators deplete Li more efficiently, additional mechanisms, such as disk locking during the protostellar phase, may be required to reconcile theory with observations. Disk locking could introduce a two-stage braking process, regulating early angular momentum loss and Li evolution while potentially mitigating magnetic field overestimates. However, disk locking remains empirically inferred and reliant on free parameters; a physically grounded formulation tied to fundamental stellar properties is still needed.\par

By incorporating time-dependent variability in both the magnetic field and convective efficiency parameter, our approach reduces the number of free parameters in stellar modeling, allowing focus on other theoretical challenges, such as magnetic field geometry, alternative dynamo mechanisms, or protostellar disk coupling. Our results underscore the potential of physics-driven refinements in stellar evolution theory, paving the way for more predictive models of Li depletion and rotational evolution.\par

\section*{Acknowledgements}
We wish to acknowledge the generous work and help offered by the MESA and TOPCAT communities. The authors acknowledge funding support from Spanish public funds (including FEDER funds) for research under project PID2023-149439NB-C43 funded by MICIU/AEI/10.13039/501100011033 and by “ERDF A way of making Europe”. AGH also acknowledges the funding support of the 'European Regional Development Fund / Andaluc\'ia-Consejera de Econom\'ia y Conocimiento' under project E-FQM-041-UGR18 by the Universidad de Granada.

\section*{Software}
The simulations used in this paper have been executed with the MESA release 10398 \footnote{\url{http://mesa.sourceforge.net/release/2018/03/21/r10398.html}}. The different figures have been generated using GNU Octave 5.1.0. \footnote{\url{https://www.gnu.org/software/octave/}}.  Gaia and GES data sets have been analyzed using TOPCAT 4.8-7 \footnote{\url{https://www.g-vo.org/topcat/topcat/}}. 

\section*{Data Availability}
The data underlying this article will be shared upon reasonable request to the corresponding author.




\bibliographystyle{mnras}
\bibliography{mblithium}



\appendix
\section{Full list of OC's}

\begin{table*}
	\centering
	\begin{tabular}{|l l l l l || c c c c c | c c c c c|} 
		\hline
             & & & & & & & $\omegaini$ & & & & & $\omegaini$ & & \\
		Open Cluster & GES\_FLD & Age (Ga) & [Fe/H] & $N_*$ & 0.095 & 0.10 & 0.105 & 0.11 & 0.115 & 0.12 & 0.125 & 0.13 & 0.14 & 0.1425\\
		\hline
            25 Ori & 25\_Ori & 0.013 & 0 & 294 & 0 & 0 & 0 & 0 & 0 & 0 & 0 & 0 & 0 & 0\\
            ASCC 50 & Assc50 & 0.011 & 0.02 & 1225 & 0 & 0 & 0 & 0 & 0 & 0 & 0 & 0 & 0 & 0\\
            \rowcolor{lightgray}
            Berkeley 21 & Br21 & 2.138 & -0.21 & 744 & 0 & 0 & 0 & 0 & 0 & 1 & 1 & 1 & 1 & 1\\
            Berkeley 22 & Br22 & 2.455 & -0.26 & 395 & 0 & 0 & 0 & 0 & 0 & 0 & 0 & 0 & 0 & 0\\
            Berkeley 25 & Br25 & 2.455 & -0.25 & 81 & 0 & 0 & 0 & 0 & 0 & 0 & 0 & 0 & 0 & 0\\
            Berkeley 30 & Br30 & 0.295 & -0.13 & 332 & 0 & 0 & 0 & 0 & 0 & 0 & 0 & 0 & 0 & 0\\
            Berkeley 31 & Br31 & 2.818 & -0.29 & 616 & 0 & 0 & 0 & 0 & 0 & 0 & 0 & 0 & 0 & 0\\
            Berkeley 32 & Br32 & 4.898 & -0.31 & 438 & 0 & 0 & 0 & 0 & 0 & 0 & 0 & 0 & 0 & 0\\
            Berkeley 36 & Br36 & 6.761 & -0.15 & 739 & 0 & 0 & 0 & 0 & 0 & 0 & 0 & 0 & 0 & 0\\
            \rowcolor{lightgray}
            Berkeley 39 & Br39 & 5.623 & -0.14 & 899 & 1 & 1 & 1 & 1 & 1 & 1 & 2 & 2 & 2 & 2\\
            Berkeley 44 & Br44 & 1.445 & 0.22 & 93 & 0 & 0 & 0 & 0 & 0 & 0 & 0 & 0 & 0 & 0\\
            Berkeley 73 & Br73 & 1.413 & -0.26 & 77 & 0 & 0 & 0 & 0 & 0 & 0 & 0 & 0 & 0 & 0\\
            Berkeley 75 & Br75 & 1.738 & -0.34 & 75 & 0 & 0 & 0 & 0 & 0 & 0 & 0 & 0 & 0 & 0\\
            Berkeley 81 & Br81 & 1.148 & 0.22 & 279 & 0 & 0 & 0 & 0 & 0 & 0 & 0 & 0 & 0 & 0\\
            Blanco 1 & Blanco1 & 0.105 & -0.03 & 463 & 0 & 0 & 0 & 0 & 0 & 0 & 0 & 0 & 0 & 0\\
            Chamaeleon I & Cha\_I & 0.001 & -0.03 & 709 & 0 & 0 & 0 & 0 & 0 & 0 & 0 & 0 & 0 & 0\\
            Collinder 197 & Col197 & 0.014 & 0.03 & 409 & 0 & 0 & 0 & 0 & 0 & 0 & 0 & 0 & 0 & 0\\
            Collinder 69 & lam\_Ori & 0.013 & -0.09 & 836 & 0 & 0 & 0 & 0 & 0 & 0 & 0 & 0 & 0 & 0\\
            Czernik 24 & Cz24 & 2.692 & -0.11 & 346 & 0 & 0 & 0 & 0 & 0 & 0 & 0 & 0 & 0 & 0\\
            Czernik 30 & Cz30 & 2.884 & -0.31 & 226 & 0 & 0 & 0 & 0 & 0 & 0 & 0 & 0 & 0 & 0\\
            ESO 92-05 & ESO92\_05 & 4.467 & -0.29 & 212 & 0 & 0 & 0 & 0 & 0 & 0 & 0 & 0 & 0 & 0\\
            Haffner 10 & Haf10 & 3.802 & -0.1 & 562 & 0 & 0 & 0 & 0 & 0 & 0 & 0 & 0 & 0 & 0\\
            IC 2391 & IC2391 & 0.029 & -0.06 & 434 & 0 & 0 & 0 & 0 & 0 & 0 & 0 & 0 & 0 & 0\\
            \rowcolor{lightgray}
            IC 2602 & IC2602 & 0.036 & -0.06 & 1836 & 0 & 0 & 0 & 0 & 0 & 0 & 1 & 1 & 0 & 0\\
            \rowcolor{lightgray}
            IC 4665 & IC4665 & 0.033 & 0.01 & 567 & 0 & 0 & 0 & 0 & 1 & 1 & 0 & 0 & 0 & 0\\
            Loden 165 & Loden165 & 3 &  & 388 & 0 & 0 & 0 & 0 & 0 & 0 & 0 & 0 & 0 & 0\\
            \rowcolor{lightgray}
            Messier 67 & M67 & 3.981 & -0.02 & 131 & 6 & 6 & 6 & 6 & 6 & 6 & 6 & 6 & 6 & 6\\
            \rowcolor{lightgray}
            NGC 2141 & NGC2141 & 1.862 & -0.04 & 853 & 1 & 1 & 1 & 1 & 1 & 1 & 1 & 1 & 1 & 1\\
            NGC 2158 & NGC2158 & 1.549 & -0.15 & 616 & 0 & 0 & 0 & 0 & 0 & 0 & 0 & 0 & 0 & 0\\
            NGC 2232 & NGC2232 & 0.018 & -0.03 & 1866 & 0 & 0 & 0 & 0 & 0 & 0 & 0 & 0 & 0 & 0\\
            NGC 2243 & NGC2243 & 4.365 & -0.45 & 661 & 0 & 0 & 0 & 0 & 0 & 0 & 0 & 0 & 0 & 0\\
            NGC 2244 & NGC2244 & 0.004 & -0.04 & 452 & 0 & 0 & 0 & 0 & 0 & 0 & 0 & 0 & 0 & 0\\
            NGC 2264 & NGC2264 & 0.003 & -0.1 & 1857 & 0 & 0 & 0 & 0 & 0 & 0 & 0 & 0 & 0 & 0\\
            \rowcolor{lightgray}
            NGC 2355 & NGC2355 & 1 & -0.13 & 208 & 1 & 1 & 1 & 1 & 1 & 1 & 1 & 1 & 1 & 1\\
            \rowcolor{lightgray}
            NGC 2420 & NGC2420 & 1.698 & -0.15 & 562 & 1 & 1 & 1 & 1 & 1 & 1 & 1 & 1 & 1 & 1\\
            \rowcolor{lightgray}
            NGC 2425 & NGC2425 & 2.399 & -0.13 & 528 & 1 & 1 & 1 & 1 & 1 & 1 & 1 & 1 & 1 & 1\\
            \rowcolor{lightgray}
            NGC 2451 & NGC2451 & 0.035 & -0.08 & 1656 & 0 & 0 & 0 & 0 & 0 & 1 & 1 & 0 & 2 & 1\\
            \rowcolor{lightgray}
            NGC 2516 & NGC2516 & 0.24 & -0.04 & 759 & 0 & 0 & 0 & 1 & 1 & 1 & 3 & 4 & 4 & 3\\
            NGC 2547 & NGC2457 & 0.032 & -0.03 & 477 & 0 & 0 & 0 & 0 & 0 & 0 & 0 & 0 & 0 & 0\\
            NGC 3293 & NGC3293 & 0.01 & -0.08 & 584 & 0 & 0 & 0 & 0 & 0 & 0 & 0 & 0 & 0 & 0\\
            \rowcolor{lightgray}
            NGC 3532 & NGC3532 & 0.398 & -0.01 & 1145 & 1 & 1 & 0 & 1 & 1 & 1 & 1 & 1 & 0 & 1\\
            NGC 3766 & NGC3766 & 0.023 & -0.12 & 399 & 0 & 0 & 0 & 0 & 0 & 0 & 0 & 0 & 0 & 0\\
            NGC 4815 & NGC4815 & 0.372 & 0.08 & 218 & 0 & 0 & 0 & 0 & 0 & 0 & 0 & 0 & 0 & 0\\
            \rowcolor{lightgray}
            NGC 6005 & NGC6005 & 1.259 & 0.22 & 560 & 0 & 0 & 0 & 0 & 0 & 0 & 0 & 1 & 1 & 1\\
            NGC 6067 & NGC6067 & 0.126 & 0.03 & 780 & 0 & 0 & 0 & 0 & 0 & 0 & 0 & 0 & 0 & 0\\
            NGC 6253 & NGC6253 & 3.246 & 0.33 & 646 & 0 & 0 & 0 & 0 & 0 & 0 & 0 & 0 & 0 & 0\\
            \rowcolor{lightgray}
            NGC 6259 & NGC6259 & 0.269 & 0.18 & 494 & 0 & 0 & 0 & 0 & 0 & 0 & 0 & 0 & 0 & 1\\
            \rowcolor{lightgray}
            NGC 6281 & NGC6281 & 0.513 & -0.04 & 320 & 0 & 0 & 0 & 0 & 0 & 0 & 0 & 0 & 1 & 1\\
            \rowcolor{lightgray}
            NGC 6405 & NGC6405 & 0.035 & -0.02 & 701 & 1 & 1 & 1 & 0 & 1 & 3 & 2 & 0 & 0 & 0\\
            NGC 6530 & NGC6530 & 0.002 & -0.02 & 1983 & 0 & 0 & 0 & 0 & 0 & 0 & 0 & 0 & 0 & 0\\
            \rowcolor{lightgray}
            NGC 6633 & NGC6633 & 0.692 & -0.03 & 1662 & 2 & 2 & 2 & 1 & 1 & 0 & 0 & 0 & 1 & 1\\
            NGC 6649 & NGC6649 & 0.071 & -0.08 & 283 & 0 & 0 & 0 & 0 & 0 & 0 & 0 & 0 & 0 & 0\\
            NGC 6705 & NGC6705 & 0.309 & 0.03 & 1066 & 0 & 0 & 0 & 0 & 0 & 0 & 0 & 0 & 0 & 0\\
            \rowcolor{lightgray}
            NGC 6709 & NGC6709 & 0.191 & -0.02 & 730 & 2 & 1 & 0 & 0 & 0 & 0 & 0 & 0 & 1 & 1\\
            NGC 6802 & NGC6802 & 0.661 & 0.14 & 197 & 0 & 0 & 0 & 0 & 0 & 0 & 0 & 0 & 0 & 0\\
            Pismis 15 & Pismis15 & 0.871 & 0.02 & 333 & 0 & 0 & 0 & 0 & 0 & 0 & 0 & 0 & 0 & 0\\
            Pismis 18 & Pismis18 & 0.575 & 0.14 & 142 & 0 & 0 & 0 & 0 & 0 & 0 & 0 & 0 & 0 & 0\\
            Ruprecht 134 & Rup134 & 1.66 & 0.27 & 680 & 0 & 0 & 0 & 0 & 0 & 0 & 0 & 0 & 0 & 0\\
            Trumpler 14 & Trumpler14 & 0.003 & -0.01 & 1902 & 0 & 0 & 0 & 0 & 0 & 0 & 0 & 0 & 0 & 0\\
            \rowcolor{lightgray}
            Trumpler 20 & Trumpler20 & 1.862 & 0.13 & 1213 & 3 & 3 & 3 & 2 & 2 & 2 & 2 & 2 & 2 & 2\\
            Trumpler 23 & Trumpler23 & 0.708 & 0.2 & 165 & 0 & 0 & 0 & 0 & 0 & 0 & 0 & 0 & 0 & 0\\
            Trumpler 5 & Trumpler5 & 4.266 & -0.35 & 1138 & 0 & 0 & 0 & 0 & 0 & 0 & 0 & 0 & 0 & 0\\
            Gamma Vel & gamma2\_vel & 0.02 & -0.02 & 1269 & 0 & 0 & 0 & 0 & 0 & 0 & 0 & 0 & 0 & 0\\
            Rho Oph & Rho\_Oph & 0.001 & 0.03 & 311 & 0 & 0 & 0 & 0 & 0 & 0 & 0 & 0 & 0 & 0\\           
            \hline
	\end{tabular}
 	\caption{List of selected OCs. For each OC, name, GES denomination, estimated age, metallicity and number of components are listed. Two sets of simulations for different $\omegaini$ are shown, where $\omegaini = \oomegac$, $\Omega$ is the star angular velocity at stellar surface, and $\omegac$ is the surface velocity at the equator of a rotating star where the centrifugal force balances the Newtonian gravity. Those OCs with member that have been selected in any of the simulations are highlighted in grey.}
  	\label{tab:oc_full_list}
\end{table*}

\clearpage
\section{Additional models visualization}
The following are a set of additional figures for 1 $\msun$ models for which the time evolution of surface \isotope[7]{Li} abundance relative to \isotope[1]{H}, $\amlt$ and angular velocities are shown.
\begin{figure*}
	\includegraphics[clip, width=\columnwidth]{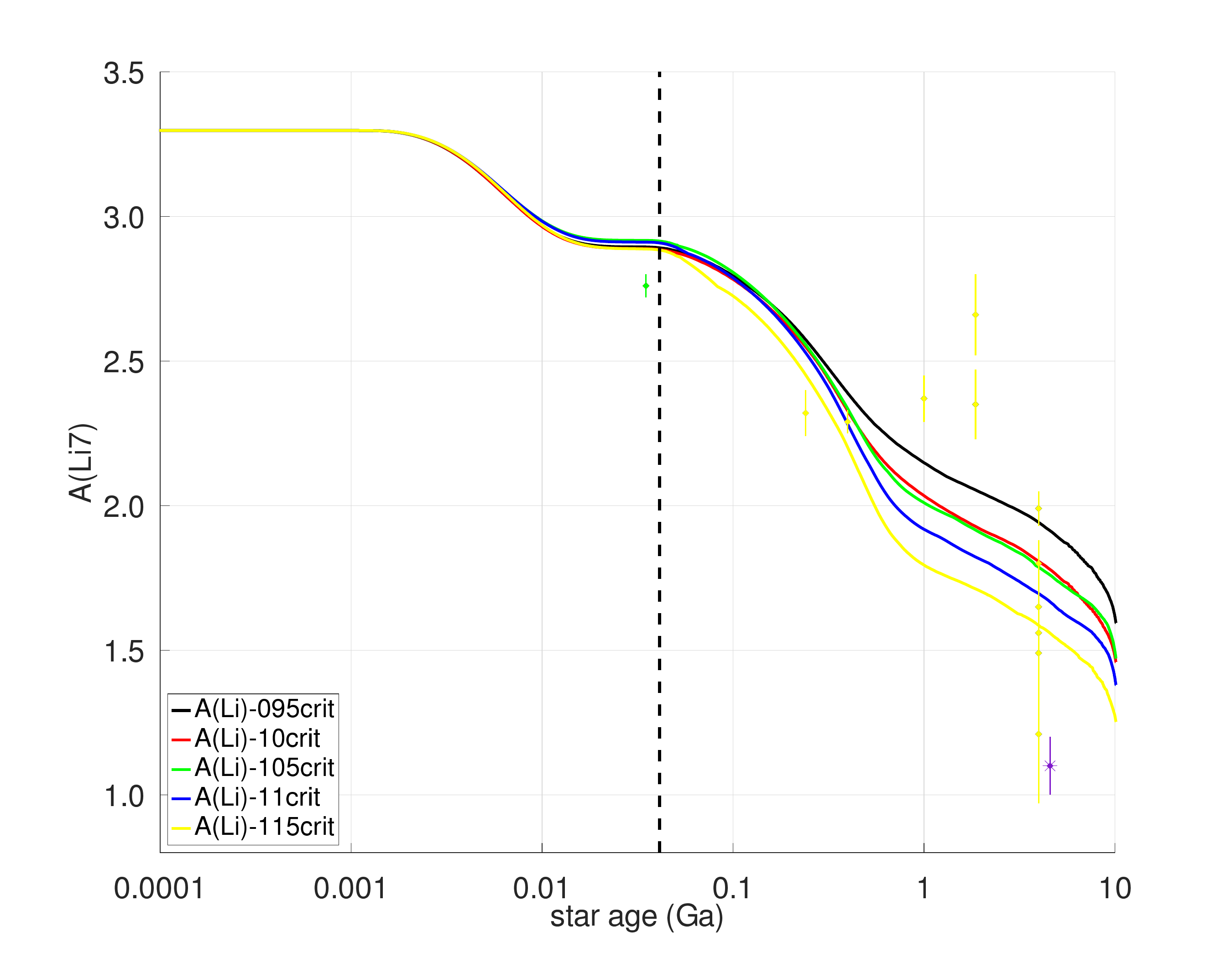}
    \caption{Ratio of surface \isotope[7]{Li} abundance to \isotope[1]{H} as a function of time is depicted for various 1 $\msun$ models with magnetic field. The models have initial rotation rates ranging from 0.095 to 0.115. The surface Li abundance for the present-day Sun (purple start) is based on the study by \citet{Asplund2009}. The other colored dots represent surface \isotope[7]{Li} abundances for stars with parameters within the specified selection intervals, corresponding to the evolution curve of the same color. The dashed vertical line indicates the Zero Age Main Sequence (ZAMS).}
    \label{fig:li_var_vel_var_g_1}
\end{figure*}

\begin{figure*}
	\includegraphics[clip, width=\columnwidth]{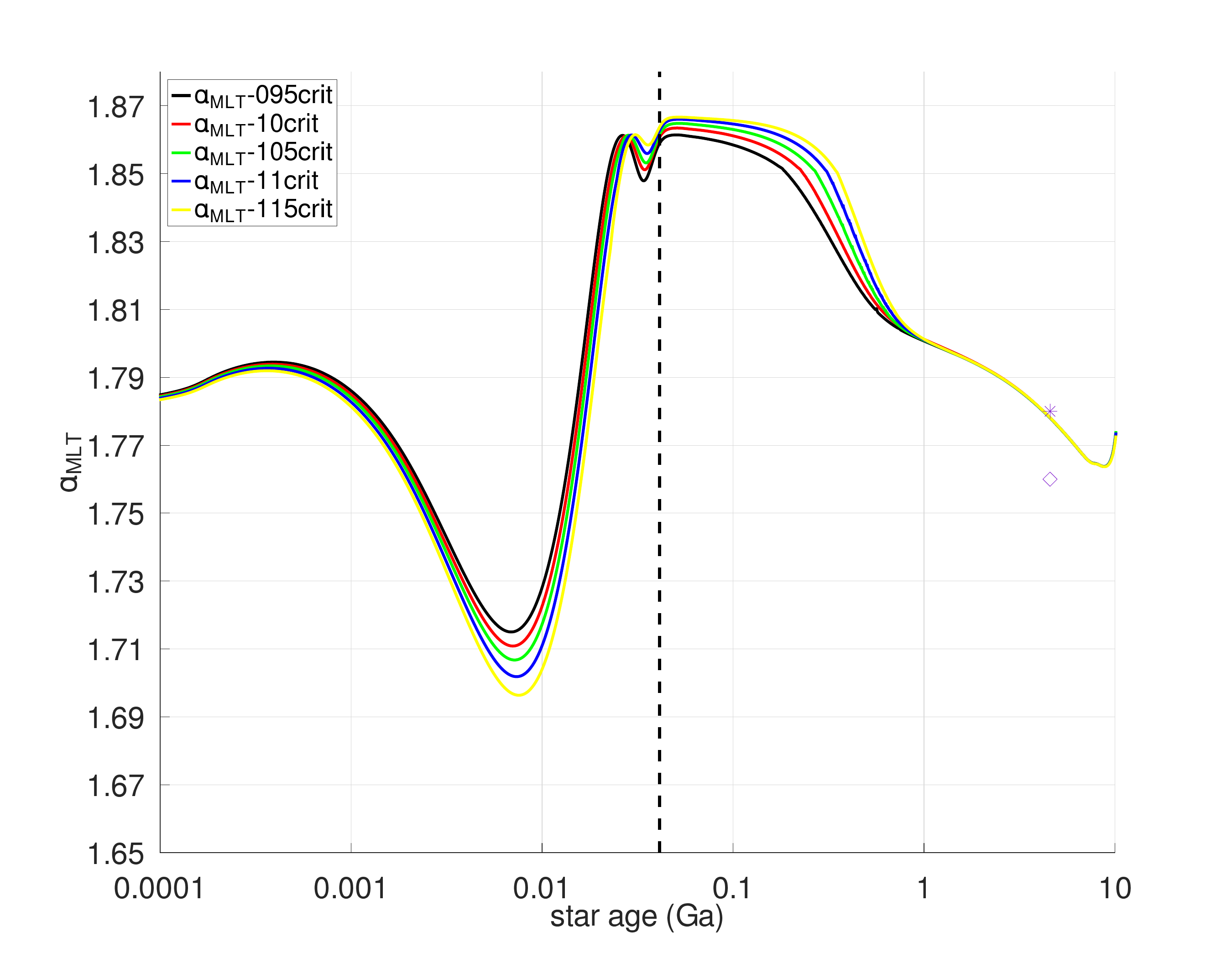}
\caption{The evolution of $\amlt$, as a function of time and $\omegaini$ for several 1 $\msun$ models. The models include initial rotation with $\omegaini$ between 0.095 and 0.115. The dashed vertical line makes reference to the ZAMS. The purple star and diamond are the $\amlt$ given by \citet{Sonoi2018} and \citet{Samadi2005}, respectively.}
\label{fig:alpha_mlt_var_vel_g1}
\end{figure*}

\begin{figure*}
    \centering
    \begin{subfigure}[h]{0.47\textwidth}
    \includegraphics[clip,width=\textwidth]{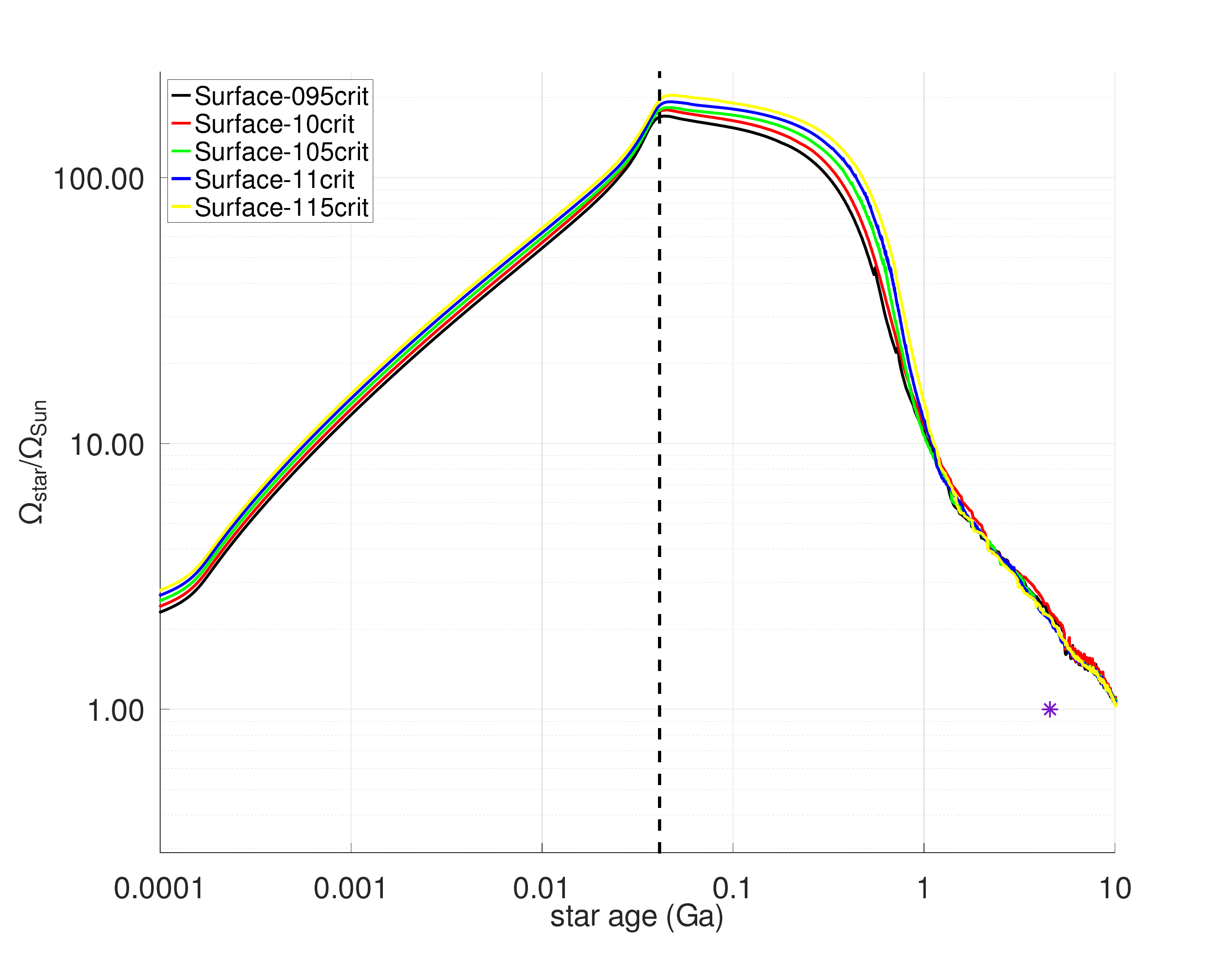}
    \label{fig:subim21}
    \end{subfigure}
    \begin{subfigure}[h]{0.47\textwidth}
    \includegraphics[clip,width=\textwidth]{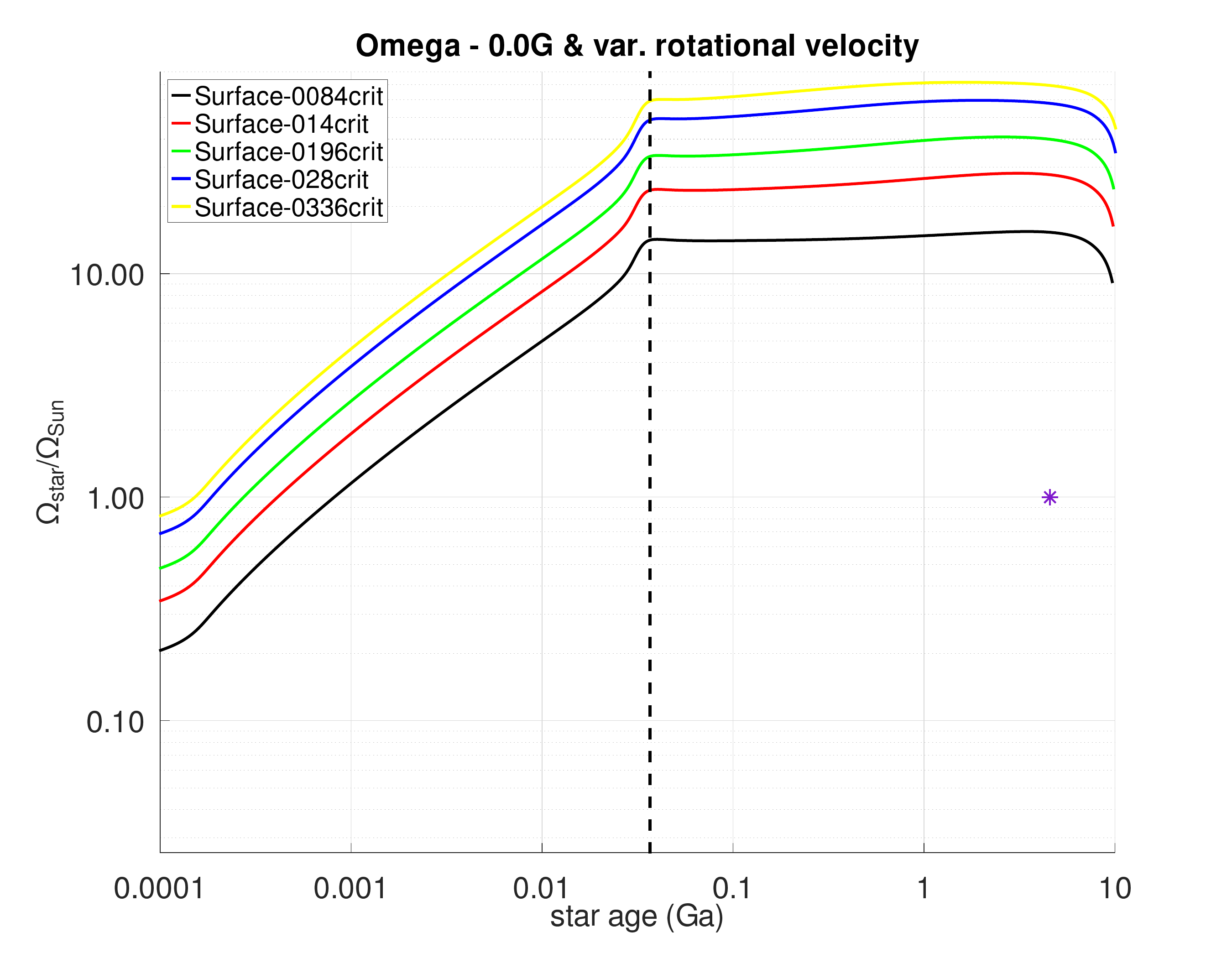}
    \label{fig:subim22}
    \end{subfigure}    
\caption{Grid showing of the evolution of surface rotational velocity, as a function of time for several 1 $\msun$ models. The left figure show a set of models with variable magnetic field with intensity, and $\omegaini$ varying between 0.095 and 0.115. In the right one the magnetic field intensity was set to $0.0\,\Gauss$, and $\omegaini$ varying between 0.0 and 0.0336. The purple star and square are surface Li abundance for the present-day Sun \citep{Asplund2009}. The dashed vertical line makes reference to the ZAMS.}
\label{fig:grid_li_var_g}
\end{figure*}

\appendix

\bsp	
\label{lastpage}
\end{document}